\documentclass{article}

\usepackage{PRIMEarxiv}

\usepackage{braket}
\usepackage[utf8]{inputenc} 
\usepackage[T1]{fontenc}    
\usepackage{hyperref}       
\usepackage{url}            
\usepackage{booktabs}       
\usepackage{amsfonts}       
\usepackage{bm}
\usepackage{nicefrac}       
\usepackage{subfig}         
\usepackage{microtype}      
\usepackage{lipsum}
\usepackage{amsmath} 
\usepackage{fancyhdr}       
\usepackage{graphicx}       
\graphicspath{{media/}}     
\usepackage{comment}
\usepackage{xcolor}

\title{Real-Time Monitoring of MHD Liquid Metal Flows with Shallow Recurrent Decoders
}

\author{
  Claudio Scardino$^a$, Stefano Riva$^{b,a}$, Carolina Introini$^a$, Matteo Lo Verso$^a$, Eric Cervi$^c$, Antonio Cammi$^{d,a}$, \\ \textbf{Laura Savoldi}$^e$ \\
  $^a$Energy Department, CeSNEF - Nuclear Engineering Division, Politecnico di Milano, Milano \\
  $^b$Autodesk Research, 6 Agar Street, London, WC2R 0QE, UK \\
  $^c$Argonne National Laboratory, Lemont, IL, 60439, United States of America \\
  $^d$Department of Mechanical and Nuclear Engineering, Emirates Nuclear Technology
  Center, Khalifa University, \\ Abu Dhabi, 127788, United Arab Emirates \\
  $^e$MAHTEP Group, Dipartimento Energia “Galileo Ferraris”, Politecnico di Torino, Torino \\
  \texttt{claudio.scardino@mail.polimi.it, stefano.riva@polimi.it, carolina.introini@polimi.it,}\\ \texttt{matteo.loverso@polimi.it, ecervi@anl.gov, antonio.cammi@polimi.it, laura.savoldi@polito.it} 
}

\begin{document}
\maketitle

\begin{abstract}
State estimation in magnetohydrodynamic flows is critical for real-time monitoring of liquid metal blankets in tokamak fusion reactors. Due to the multiphysics nature of these phenomena, high-fidelity simulations are computationally prohibitive for real-time applications. This work investigates a data-driven Reduced Order Model framework: the Shallow Recurrent Decoder (SHRED) coupled with Principal Component Analysis, to map sparse temperature measurements to the full thermo-hydraulic system's state. The major contribution of this work lies in the two-parameter analysis of a fully three-dimensional domain representative of the DEMO breeding blanket configuration. Here, the flow is subjected to an external magnetic field varying in direction and intensity and is hindered by two cylinders acting as a water-cooling system, which impose a temperature boundary condition on their surfaces. This double-parametric magnetic variation induces nonlinear transitions in the flow dynamics, ranging from chaotic behavior at low magnetic field intensities to laminarized regimes at high intensities, characterized by the formation of asymmetric side layers at an inclination angle of 30 degrees. SHRED reconstruction maintains a mean relative error of approximately 5\% for the temperature, pressure, and velocity fields. This accuracy is maintained across both weak and strong magnetic fields, ranging from 0.075 T to 0.300 T, and for inclination angles from 5 to 30 degrees, reflecting its dominant toroidal component. These errors are only slightly larger than the lower error bound dictated by low-rank truncation. The results establish SHRED as a reliable state estimator for complex and realistic engineering applications involving completely unseen parametric scenarios and validate it as an accurate real-time state estimation technique suitable for online monitoring and control of real facilities.
\end{abstract}


\section{Introduction}

Magneto-hydrodynamics (MHD) \cite{RevModPhys.76.1071} describes the coupling between electromagnetic fields (specifically, the electric field $\mathbf{E} \ (\text{V/m})$ and the magnetic field $\mathbf{B}$ ($\text{T}$)) and thermal fluid dynamics fields, such as velocity $\mathbf{u}$ ($\text{m/s}$), temperature $T$ ($\text{K}$), and pressure $p$ ($\text{Pa}$), for an electrically conductive fluid, as in the case of liquid metals \cite{buhler2007liquid}. In a Tokamak fusion reactor \cite{wesson2011tokamaks}, the magnetic confinement of the plasma, achieved by exploiting intense external magnetic fields $\mathbf{B}_{\text{ext}}$, directly influences the fluid dynamics of the liquid metal flowing in the breeding blanket, thereby altering its fluid-dynamics state. Knowledge of the liquid metal behavior inside the blanket under different external magnetic fields is of primary concern, as it dictates the heat extraction from the plasma and, subsequently, the heat exchange with the water cooling circuit of the reactor. In fact,  in an electrically conducting fluid, the overall magnetic field (given by the superposition of the external one and of the internally generated one due to charge motion), influences the thermal-hydraulic state, acting on the fluid flow which, in turns, influences the internal magnetic field. As a result, a continuous two-way coupling is established between the electromagnetic field (described by the Maxwell equations) and the fluid dynamics (described by the Navier-Stokes equations), resulting in a highly non-linear system of coupled equations within a multiphysics framework \cite{biskamp1997nonlinear}. 


The inherent multi-physics and non-linear nature of the  MHD equations leads to high computational cost when solving the high fidelity system and tracking the MHD state; indeed, even for the same geometry and identical initial and boundary conditions, problems that differ only in some parameter values, such as the intensity of the magnetic field or its inclination angle, can produce very different solutions: for example, intense magnetic fields induce a laminarization of the flow, as the Lorentz forces act as a brake, suppressing three-dimensional turbulent structures, implying a very different dynamics between high and low magnetic fields. Consequently, in order to identify the overall dynamics of the flow, for any combination of the external parameters, a large number of high-fidelity simulations are needed; however, this is not feasible in industrial applications due to the high computational and time costs. In particular, with a view to real-time applications such as control and monitoring, this computational burden is excessive, making online state estimation unfeasible, as simulations are required near-instantaneously.

In this context, over the last few years, data-driven reduced-order techniques have established themselves as reliable and accurate state estimators in the nuclear sector, primarily with reference to fission reactors \cite{riva_multi-physics_2024,riva_hybrid_2023-1,riva2025_parametricMSFR,Gong03062022,GONG2022109431}. Conversely, their application to MHD problems remains highly limited to plasma applications (plasma control \cite{Degrave}, instability mitigation \cite{Seo}, and profile regulation \cite{Jalalvand} in tokamaks), and even more so when considering liquid metal blankets. 

Alongside reduced-order approaches, Deep Learning methods have also been investigated in fission-reactor applications, including CNN-based reconstruction of multiphysics reactor fields from sparse and moving sensors \cite{gong_reactor_2024}, data-enabled physics-informed neural networks for neutron-diffusion eigenvalue problems \cite{YANG2023109656}, and physics-informed CNNs for temperature-field monitoring in high-temperature gas reactors \cite{leite_application_2025}.

Nevertheless, when dealing with the extremely high dimensionality of 3D full-order simulations, according to the size of the mesh ($\mathcal{N}_h \sim\mathcal{O}(10^5 - 10^8)$), standard Deep Learning architectures, such as deep Convolutional Neural Networks (CNNs), involve up to tens of millions of trainable parameters (the so-called curse of dimensionality) \cite{brunton_data-driven_2022}. This vast number of parameters makes these architectures overly complex, memory-intensive, slow to train, and inherently data-inefficient. Indeed, to prevent overfitting, such large networks require an impractically massive dataset of Full Order Model (FOM) simulations for training: given the cost of multi-query simulations for coupled, non-linear problems, generating a dataset with a sufficient number of training cases becomes fundamentally unfeasible.

A promising alternative is the integration of shallow architectures trained on a latent representation of the high-fidelity simulations within a reduced mathematical space. Specifically, the problem dimensionality can be compressed using Reduced Order Modeling (ROM) techniques \cite{brunton_data-driven_2022}, such as the Principal Component Analysis (PCA) adopted in this work. This approach allows extracting the dominant spatial patterns, referred to as basis functions $\{\phi_i\}_{i=1}^r$, recurrent across various flow regimes induced by the externally applied magnetic field. These functions are then exploited to build a linear reduced space defined by their span: $X_r=\text{span}(\{\phi_i\}_{i=1}^r)$, where $r$ is the truncation rank and hence the dimensionality of the surrogate model. In this latent space, each snapshot of a particular spatial field $\psi$ (obtained for a given external magnetic field $\mathbf{B}_{\text{ext}}$ at a given time $t$) is represented through its projection as a linear combination of the basis functions and the latent coefficients $\{\alpha_i(t;\mathbf{B}_{\text{ext}})\}_{i=1}^r$, which encode the latent dynamics. For the reduction step, this work adopts the Principal Component Analysis (PCA), rather than the un-centered Singular Value Decomposition commonly used in fluid dynamics, to emphasize the neural network's capability of learning the physical oscillations around the mean field. By subtracting the mean snapshot $\overline{\psi}(\mathbf{x})$, the network avoids wasting parameters on capturing the dominant static mean field, as would otherwise be required with standard SVD.


In a Data Assimilation context involving physical facilities, it is natural to link sensor measurements, rather than the system parameters, directly to the latent dynamics, as the parameters may be a priori unknown. Because SHRED works directly with sparse measurements rather than parameter values, it is inherently suitable for direct deployment in engineering facilities. In this scenario, once the basis functions are extracted from the training snapshots, the neural network only needs to learn the non-linear mapping between the sparse measurements and the latent dynamics. During the online phase, as new measurements become available, the corresponding latent coefficients are rapidly inferred by the network. Finally, these coefficients are linearly combined with the previously computed basis functions to perform the full state estimation. As the dimensionality reduction is performed via PCA, at a given time $t$ for a new value of the magnetic field $\mathbf{B}_{\text{ext}}^\star$, the neural network must determine only $r \sim \mathcal{O}(10^1)$ unknowns, rather than the $\mathcal{N}_h \sim \mathcal{O}(10^5 - 10^8)$ unknowns dictated by the size of the mesh. This implies that a Deep Neural Network is no longer required, as a shallow decoder with fewer than 200'000 parameters is sufficient. By adopting a shallow architecture, the network becomes intrinsically fast to train and data-efficient; the reduced number of parameters mitigates the overfitting problem, thereby eliminating the need for a massive training dataset and drastically reducing the computational cost associated with data generation. Moreover, to naturally handle the temporal structure of the input extracted from sparse sensors, a Recurrent Neural Network, such as a Long Short-Term Memory (LSTM) network \cite{hochreiter1997long}, can be adopted to map the measurements into an intermediate latent space. By coupling the LSTM with a shallow decoder \cite{erichson_shallow_2020}, which takes the intermediate representation as input and outputs the latent dynamics $\{\alpha_i(t;\mathbf{B}_{\text{ext}}^\star)\}_{i=1}^r$, the SHallow REcurrent Decoder (SHRED) architecture is built. This is the specific framework adopted in the present work.

Relying on specific input measurements, SHRED is able to perform indirect state estimation; by processing data from only a single scalar field, such as the temperature field adopted in this work, the model is capable of reconstructing all physical fields, including the unmeasured ones. This is a crucial feature since, in an engineering-oriented framework, not all variables are equally accessible. For instance, acquiring localized measurements for vector fields, such as velocity, is practically unfeasible. Moreover, when SHRED is trained on noisy measurements to target the latent dynamics, during offline training the network learns to effectively decouple the noise from the pure, uncorrupted measurements. 

Despite the capability of SHRED to perform state estimation by handling the strong non-linearities between sensor measurements and latent dynamics, inherent to multiphysics problems, its adoption in the fusion field \cite{kutz_shallow_2024} is not yet as widespread as its counterpart applications in fission reactors \cite{riva2024robuststateestimationpartial,riva2025towards,riva2025constrainedsensingreliablestate,From_Models_To-Experiments} or other physics fields \cite{williams_sensing_2023,tomasetto2025reducedordermodelingshallow}. Nevertheless, recent investigations \cite{loverso2025reduced} demonstrate the feasibility of this architecture to perform state estimation for a two-dimensional MHD flow using SHRED. It should be mentioned that a first preliminary study on a three-dimensional benchmark case, estimating the MHD state of Pb-Li around a cooling pipe inside a blanket elementary cell, was conducted in \cite{verso2026applicationparametricshallowrecurrent}. In their work, both the intensity and the orientation of the magnetic field were simultaneously considered as parameters. While this study provided valuable insights, highlighting the potential of SHRED for accurate state reconstruction under general magnetic field conditions, further investigations involving more complex geometric configurations are required. Indeed, their benchmark considered a simplified setup focusing exclusively on the MHD flow interaction with the pipe, treating the external duct walls as virtual, open boundaries that the fluid could cross. In addition, compared to previous works, the present one focuses more on the physical interpretation of the network output.

Conversely, this work simulates the flow of liquid Pb-Li within a three-dimensional domain representing a simplified elementary cell of a breeding blanket. The liquid metal flows longitudinally inside a square duct; to model the heat exchange with the secondary coolant system, the flow is intercepted by two transversal cylinders acting as cooling pipes. The cooling process is numerically accounted for by imposing a fixed temperature boundary condition on the cylindrical surfaces, which are maintained at a lower temperature relative to the bulk liquid metal. In addition to providing a more realistic representation of a breeding blanket geometry, this benchmark case is highly interesting from a phenomenological perspective. Indeed, it allows for the characterization of distinctive MHD phenomena, such as the formation of side layers and the magnetic braking effect that suppresses the von Kármán vortex streets behind the cylinders. The selected parameters are the intensity of the external magnetic field $\mathbf{B}_{\text{ext}}$ and its inclination angle $\alpha$, which dictates the development of asymmetric side layers.

The remainder of the paper is organized as follows. Section \ref{sec: SHRED} investigates the SHRED architecture in detail, illustrating its key characteristics. In Section \ref{res}, the MHD ruling equations are presented together with the benchmark geometry, followed by the numerical results of the SHRED state estimation for the single and double parametric cases. Finally, Section \ref{concl} summarizes the most relevant conclusions and outlines future developments.

\section{The SHallow REcurrent Decoder}
\label{sec: SHRED}

The SHallow REcurrent Decoder (SHRED) is a neural network architecture, proposed by Williams \cite{williams_sensing_2023} and subsequently investigated by \cite{kutz_shallow_2024,riva2024robuststateestimationpartial,ebers_leveraging_2023,shredrom}, composed of a Long Short Term Memory (LSTM) network \cite{hochreiter1997long} and a Shallow Decoder Network (SDN) \cite{erichson_shallow_2020} that maps temporal trajectories of measurements collected from sparse (even randomly distributed) sensors of a measurable field to the full physical space. In the context of reduced order modeling, SHRED maps the temporal trajectories onto the low-dimensional latent representation of the system dynamic and hence onto the temporal evolution of the reduced coefficients. The main advantages of this architecture are:
\begin{itemize}
    \item \textbf{Natural handling of time-series}: the LSTM block is intrinsically designed to process sequential data such as measurements coming from sensors, making it the ideal tool to learn the relations with the latent dynamics; it can also handle  noisy measurements if trained to map them to the reduced coefficients. 
    \item \textbf{Extreme sensor sparsity and robustness}: thanks to temporal encoding, the network can reconstruct the full physical space using a remarkably low number of sensors (even just three in a two-dimensional domain, or four in a three-dimensional domain), while being completely agnostic to their spatial placement, allowing for ensemble training strategies \cite{riva2024robuststateestimationpartial,shredrom}.
    \item \textbf{High computational efficiency}: the combination of a compressed latent space produced by the LSTM and a shallow decoder results in very fast training times, ensures high data efficiency and requires minimal hyperparameter tuning compared to deep autoencoders \cite{williams_sensing_2023}.
\end{itemize}

Another key feature of SHRED is its ability to estimate the latent coefficients of all fields in the snapshot using measurements from a single observable field, i.e. the possibility to perform indirect state estimation. This feature is extremely useful in environments in which some operational variables are inherently easier to measure than others, such as temperature in Tokamak environments. Moreover, it has been shown by the author \cite{CS} that temperature measurements for MHD systems encode and preserve the physical causality between thermo-hydraulic  fields when processed by SHRED. Consequently, utilizing the temperature field yields a significantly lower mean relative error compared to employing alternative fields, such as pressure, for network training. Thus, throughout this work, the neural network is always trained using temperature measurements extracted from sparse sensors.

\subsection{SHRED Architecture}

In its default architecture SHRED is composed of a Recurrent Neural Network (such as the LSTM \cite{hochreiter1997long}), with two hidden layers of 64 neurons each, followed by a Shallow Decoder Network (SDN) \cite{erichson_shallow_2020}: the LSTM processes time-lagged embedded temporal measurements, characterised by the \textit{lag} hyperparameter $L$, and outputs a latent temporal representation, which is fed as input to the SDN, a simple feedforward neural network composed of two hidden layers (of respectively 350 and 400 neurons each) which maps from the latent space to the $r$ PCA coefficients at time $r$, where $r$ is the rank of truncation of the PCA for the single field.

The total number of network hyperparameters remain below 1000: despite being a shallow network, SHRED performs remarkably well in state estimation, without the need for hyperparameter tuning (e.g., changing the number of neurons and layers). This is explained by the use of a recurrent unit, such as the LSTM, which is inherently designed for time series, where time is not a simple parameter but encodes the causality and inertia of the system from which the measurements are taken. In this context, the LSTM produces a latent space that can be related to the \textit{Takens embedding} theorem \cite{10.1007/BFb0091924}. This theorem can be interpreted as the mathematical foundation of the LSTM architecture: according to it the latent space built by the LSTM well encodes the sequentiality of the data. This explains why a shallow decoder is able to well estimate the PCA coefficients, according to the \textit{Universal Approximation Theorem} \cite{HORNIK1989359,Chen1995_UniversalOperator}.

The only hyperparameter that requires tuning according to the specific application is the \textit{lag} parameter $L$. According to its definition, its value has a physical interpretation; it represents how many time steps the network has to remember to determine the state at a given time $t$, and hence it is directly linked to the characteristic time scale of the phenomenon under investigation. This is a crucial advantage, since a relationship can be found between the total time interval simulated during the offline phase and the \textit{lag} value at which the reconstruction error saturates. This ratio identifies a characteristic time scale that remains consistent across different scenarios for the same physical problem, effectively encoding the temporal dynamics (or inertia) of the system. In particular, if $L=1$, SHRED coincides with an SDN with additional hidden layers. In this sense, SHRED can be seen as a generalization of an SDN.

\subsection{SHRED as a Nonlinear Mapping to the ROM Latent Space}\label{TLE}

The full potential of the SHRED architecture emerges when coupled with dimensionality reduction techniques, since the training process becomes significantly faster, and computationally cheap, due to the substantially lower dimensionality of the output data. Since the activation functions are non-linear (e.g., sigmoid and hyperbolic tangent), SHRED effectively learns a non-linear mapping from the sensor measurements to the reduced space, thereby capturing the non-linear dependence of the PCA coefficients on the training parameter  $\alpha(\mathbf{B}_{\text{ext}})$, $\mathbf{B}_{\text{ext}} \in \Xi^{\text{train}}_{\mathbf{B}_\text{ext}}$, through sparse measurements.

The implementation of SHRED within the PCA based data driven ROM framework is now briefly summarized, covering both the offline and online stages. Its numerical implementation relies on the \textit{NuSHRED} \cite{riva2024robuststateestimationpartial,riva2025towards} library\footnote{Available at \url{https://github.com/ERMETE-Lab/NuSHRED}.}, developed by the \textbf{ERMETE Lab} at \textbf{Politecnico di Milano}. This software framework, hosted on GitHub, is built upon the original implementation by Williams et al. \cite{williams_sensing_2023}.

\begin{figure}[ht]
    \centering
    \includegraphics[width=1\linewidth]{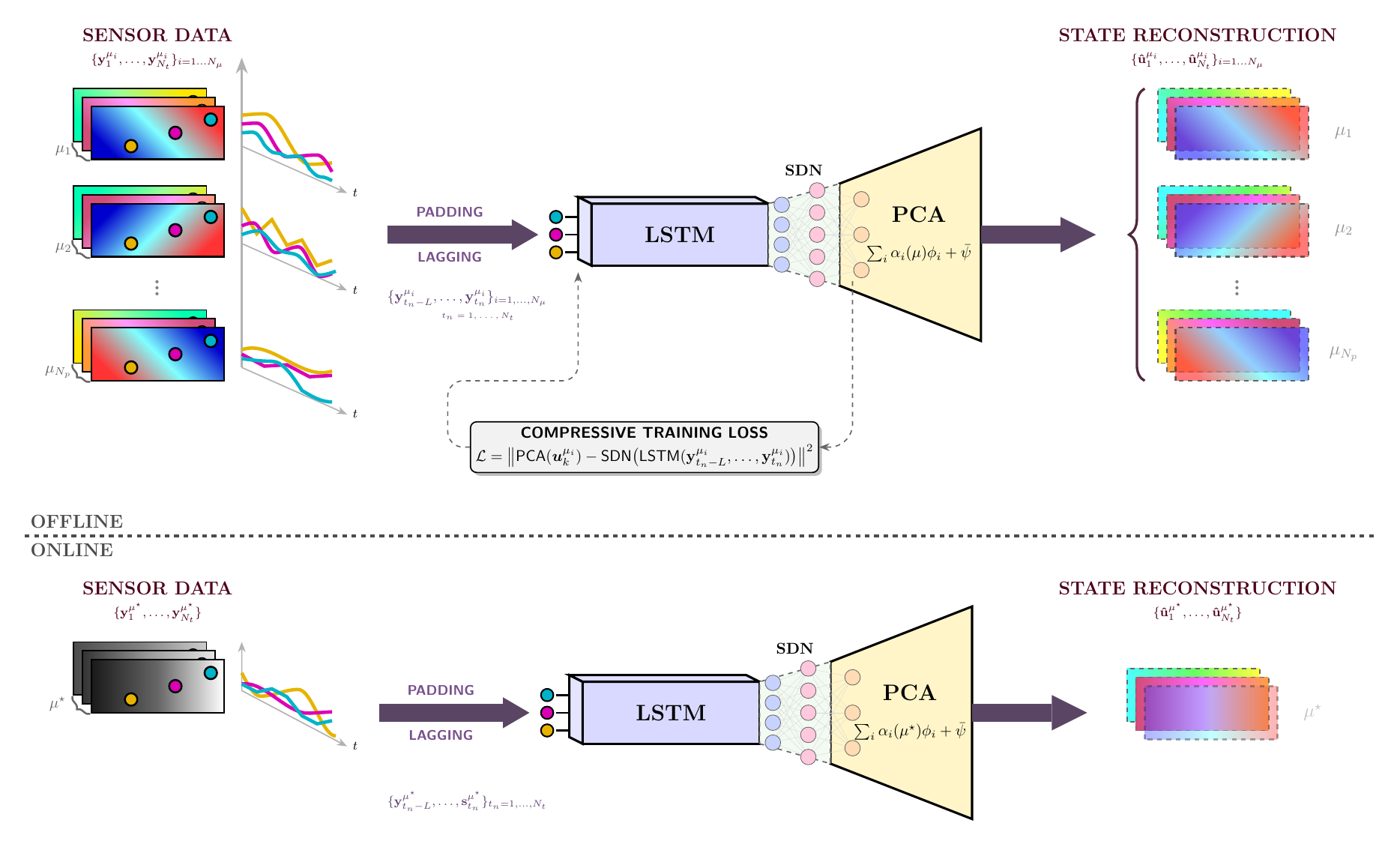}
    \caption{SHallow REcurrent Decoder architecture \& Reduced Order Modeling.}
    \label{fig:bello}
\end{figure}

A graphical representation of both offline and online stages is provided in Figure \ref{fig:bello}.
The offline training phase begins with the Principal Component Analysis of the snapshot matrices, which collect the high-fidelity solutions across all time instants and different external magnetic fields, computed separately for each physical field $\psi$. Subsequently, the temporal measurements of a single field extracted from these training snapshots are utilized to train the network. By exploiting $4$ sensors randomly placed in the physical domain, and considering $\mathcal{N}_s^{train}$ training snapshots, we aim to perform indirect reconstruction \cite{riva2024robuststateestimationpartial} and hence estimate the temporal evolution of $r$ PCA coefficients for each target field (pressure, temperature and velocity), for a total of $R$ coefficients. In the offline SHRED training stage, the input tensor contains the offline measurements $\mathbf{y}^s$, where the superscript indicates different temporal trajectories for sensors $s_1,s_2,s_3$ and $s_4$, and the targets are the offline PCA coefficients $\{\alpha_i(t)\}_{i=1}^R$. 

The input data must are pre-processed according to the Time-Lagged Embedding (TLE) procedure. In particular, for each time step $t$, a sliding window is constructed that encompasses the recent history of measurements, spanning from time $t-L$ to time $t$, for a total number of steps equal to the \textit{lag} $L$. 
Specifically, for each time instant $t_n$ lower than the value of the $lag$ ($n < L$), \textbf{pre-padding} is applied to ensure the same length of each sample: a vector containing zeros of length $L-n$ is concatenated upstream of the first non-null value: $$ \text{Input at } t_n: \quad [ \ \underbrace{0,...,0}_{L-n}, \mathbf{y}(t_0), ..., \mathbf{y}(t_n)] \quad \text{if } n < L. $$

At the end of the offline phase, SHRED has successfully learnt the non-linear relation between measurements and the reduced dynamics through the minimization of the compressive training loss function on the reduced coefficients:
$$\mathcal{L}=\| \alpha_{|\text{PCA}}- \alpha_{|\text{SDN}(\text{LSTM}(\mathbf{y}_{\text{TLE}}))}\|^2 $$
whereas $\alpha_{|\text{PCA}}$ are the latent coefficients computed through the PCA, while $\alpha_{|\text{SDN}(\text{LSTM}(\mathbf{y}_{\text{TLE}}))}$ are the ones estimated by SHRED starting from the lagged measurements.\\
In the online/test phase, starting from sparse measurements of a single field for a new parametric scenario $\mathbf{B}_{\text{ext}}^\star \in \Xi^{\text{test}}_{\mathbf{B}_\text{ext}}$, i.e., temporal trajectories at fixed points, the input tensor is prepared according to the TLE procedure. The input is processed by the trained SHRED, and the estimated PCA coefficients $\{\hat{\alpha}_i(t;\mathbf{B}_{\text{ext}}^\star)\}_{i=1}^R$, which describe the latent dynamics of all the fields, are computed. The predictive capability of SHRED in this context is not limited to temporal reconstruction but includes the ability to \textit{learn} the physical behavior within the parameter space, being able to correctly infer the dynamics of unseen parametric scenarios. Finally, the full state estimation can be performed through a linear combination of the PCA coefficients with the respective basis functions $\{\phi_i\}_{i=1}^r$ computed during the offline stage from the training snapshots:
\begin{equation}
    \hat{\psi}(\mathbf{x};t,\mathbf{B}_{\text{ext}}^\star)= \sum_{i=1}^r \phi_i(\mathbf{x}) \hat{\alpha}_i(t,\mathbf{B}_{\text{ext}}^\star) + \bar{\psi}(\mathbf{x})
\end{equation}
where $\hat{\psi}$ is the general field estimated (e.g. velocity, temperature, pressure). 

\subsection{SHRED Agnosticism to Sensor Positioning and Ensemble Strategy}
\label{ens}

SHRED has been proven to be \textit{agnostic} to sensor placement \cite{riva2025_parametricMSFR}. Agnostic means that there is no need for an algorithm for sensor positioning, provided that appreciable dynamics is present, as state estimation accuracy does not strongly depend on the measurement locations. This represents a major advantage, as sensors can be positioned based on practical accessibility and engineering constraints without compromising model performance.  Furthermore, this implies that, even if temporal trajectories coming from different positions produce slightly different state estimations, the reconstruction converges to a faithful one, considering the average PCA coefficients computed by the SHRED across different sensor placement configurations. Moreover, the shallow nature of SHRED implies an overall compact architecture which does not require time-consuming hyperparameter tuning or training. Consequently, this significantly reduces the total number of trainable parameters, leading to a remarkably fast and computationally inexpensive training phase, within the reach of even a personal laptop.

The agnosticism and the inexpensive training time combined allow the adoption of an ensemble strategy, which refers to the state estimation obtained by averaging multiple SHRED reconstructions corresponding to different sensor placement configurations. The advantage of the ensemble strategy is that the state estimation is more robust in the presence of random noise in the measurements, and moreover, an estimator of the reconstruction uncertainty can be computed. In particular, considering $K$ configurations of sensors available, the average reconstructed field $\braket{\hat{\psi}(\mathbf{x};t,\mathbf{B}_{\text{ext}}^\star)}_K$ can be computed according to:
\begin{align}
    \braket{\hat{\psi}(\mathbf{x};t,\mathbf{B}_{\text{ext}}^\star)}_K=\frac{1}{K} \sum_{k=1}^{K} \hat{\psi}_k(\mathbf{x};t,\mathbf{B}_{\text{ext}}^\star) &= \sum_{i=1}^r \left (\frac{1}{K}\sum_{k=1}^K \hat{\alpha}_{i,k}(t; \mathbf{B}_{\text{ext}}^\star)\right) \phi_i(\mathbf{x})+\bar{\psi}(\mathbf{x}) \\&=\sum_{i=1}^r\braket{\hat{\alpha}_i(t;\mathbf{B}_{\text{ext}}^\star)}_K \phi_i(\mathbf{x}) + \bar{\psi}(\mathbf{x});
\end{align}

indeed, due to the linearity of the reduced space, the average reconstruction is the linear combination of the average estimated PCA coefficients, $\braket{\hat{\alpha}_i}_K=\frac{1}{K}\sum_{k=1}^K \hat{\alpha}_{i,k}(t, \mathbf{B}_{\text{ext}}^\star)$, with the respective basis functions. Computationally speaking, it is far less expensive to compute $\braket{\hat{\psi}}_K$ through $\braket{\hat{\alpha}_i}_K$ instead of averaging all the reconstructed fields $\hat{\psi}_k$ obtained by the $k$-th configuration ($k \in \{1,...,K\}$), thanks to the dimensionality reduction of the latent space.

In order to quantify the variability of the prediction, the Empirical Ensemble Standard Deviation is computed as follows:
\begin{equation}
    \sigma^{emp}_{\psi}(\mathbf{x}; t, \mathbf{B}_{\text{ext}}^\star) = \sqrt{\frac{1}{K-1} \sum_{k=1}^{K} \left( \hat{\psi}_k(\mathbf{x}; t, \mathbf{B}_{\text{ext}}^\star) - \braket{\hat{\psi}(\mathbf{x};t,\mathbf{B}_{\text{ext}}^\star)}_K \right)^2}
    \label{emp_std}
\end{equation}

It can be noted that, in general, high spread correlates empirically with high reconstruction error. Specifically, spatial regions and time instants characterized by large values of the standard deviation indicate where and when the solution is more difficult for SHRED to accurately capture, and therefore where larger reconstruction errors are likely to occur. Moreover, in physical facilities and during real-time applications, the full-order solution is obviously not available, and hence the error committed by SHRED in estimating the thermo-hydraulic state cannot be exactly computed. The capability to estimate the accuracy of the reduced solution, both spatially and temporally, through the ensemble SHRED approach is a crucial aspect; it elevates SHRED from a mere mathematical prediction tool to a reliable, trustworthy instrument for industrial applications.

\section{Numerical Results}
\label{res}
This work studies the applicability of SHRED in a ROM context to the flow of liquid Pb-Li within a three-dimensional domain representing a simplified elementary cell of a breeding blanket for a Tokamak fusion reactor \cite{SIRIANO2025126840,MARTELLI201848}. In order to simulate the heat exchange with the secondary coolant system, the flow of the liquid metal is hindered by two transversal cylinders, which act as pipes through which the water flows, cooling the metal. The cooling process is simulated by imposing a temperature boundary condition on the cylindrical surfaces of the pipes, such that the surface is colder than the bulk temperature of the metal.

\begin{figure}[htbp]
    \centering
    \includegraphics[width=1\linewidth]{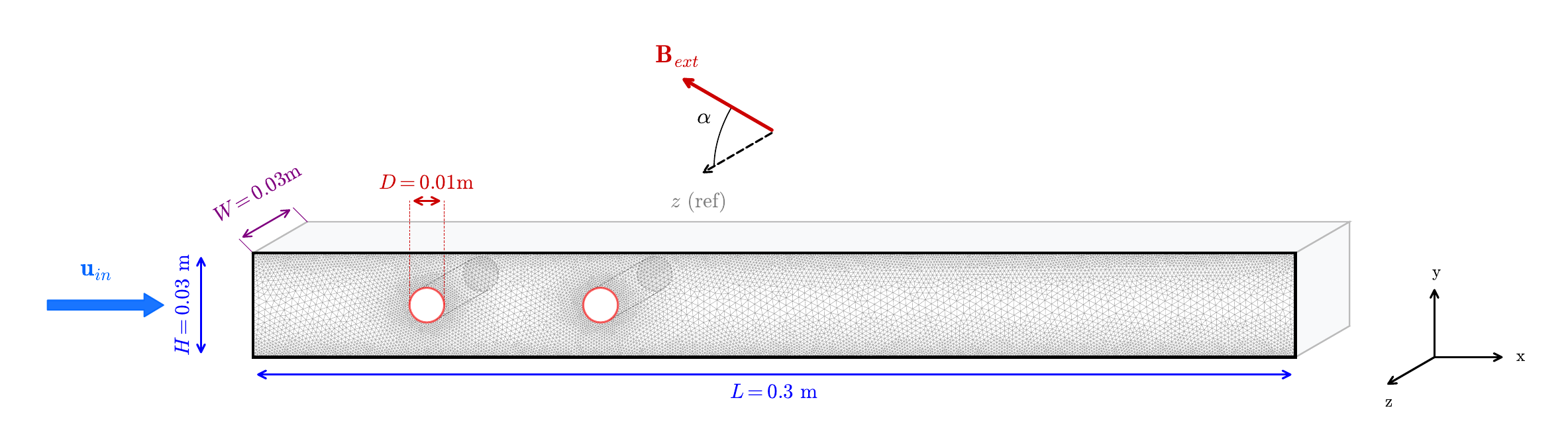}
    \caption{Simplified elementary cell of a breeding blanket: geometry and mesh, together with the representation of the applied external magnetic field.}
    \label{channel}
\end{figure}

In more detail, a three-dimensional rectangular channel with a square cross-section, with a length of 0.3 meters and a height of 0.03 meters, in which two identical cylinders of diameter $D=0.01$ meters are transversely arranged with respect to the flow, is considered. This geometry, shown in Figure \ref{channel}, is discretized using a bi-dimensional extruded mesh, consisting of 196200 cells in the 3D volume. The liquid metal flowing towards the positive $x$ direction, with inlet velocity $\mathbf{u}_{in}$, is a \textit{eutectic lead–lithium} alloy (Pb-Li) with a reference density $\rho_0 = 9806 \ \text{kg/m}^3$, dynamic viscosity $\mu = 1.93 \times 10^{-3} \ \text{Pa}\cdot\text{s}$ (for $Re \approx 10{,}000$), and Prandtl number $Pr = 0.0175$ at a reference temperature of $T_0 = 600$ K \cite{MARTELLI2019183}. The external magnetic field, with varying intensity, lies in the $z,y$ plane, with an angle of inclination $\alpha$ with respect to positive $z$ coordinate.

The full order simulations of the MHD flow have been conducted using a transient solver in \textit{OpenFOAM}, called \textbf{magnetoHDFoam} \cite{loverso2025magnetohdfoam,loverso2024novel}, for a pseudo-compressible, buoyant flow of an electrically conducting fluid, under the influence of a magnetic field, in particular using the flexible PIMPLE algorithm \cite{greenshieldsweller2022} for time-resolved and pseudo-transient simulations. The set of equations is listed Appendix \ref{A}.

To uniquely identify the MHD problem, the initial and boundary conditions must be specified. Specifically, let $\Omega$ be the entire spatial domain, $\Gamma_{\text{in}}$ the inlet, $\Gamma_{\text{out}}$ the outlet, $\Gamma_{\text{cyl}}$ the cylinder boundaries, $\Gamma_{\text{walls}}$ the walls of the square duct and $\mathbf{n}$ the outward normal vector. In this work, considering the external magnetic field oriented mainly along the $z$-axis, the upper and lower walls of the duct are defined as \textit{Side} walls, being parallel to the magnetic field, while the vertical ones are termed \textit{Hartmann} walls, being perpendicular to the magnetic field, such that $\Gamma_{\text{walls}} = \Gamma_{\text{Side}} \cup \Gamma_{\text{Hart}}$.

In this work, boundary conditions that establish a Hunt flow \cite{buhler2007liquid} are adopted. Specifically, the boundary conditions for the reduced pressure, velocity, temperature, and magnetic fields are detailed below:
\begin{align}
    \begin{cases}
        \mathbf{u}\big{ |}_{\Gamma_{\text{in}}} = |\mathbf{u}_0| \hat{x} \quad \text{with } |\mathbf{u}_0| = 0.1 \, \text{m/s}\\
        \mathbf{u}\big{|}_{{\Gamma}_{\text{walls} }\cup \Gamma_{\text{cyl}}}= \mathbf{0} \quad \text{No slip condition}\\
        \frac{\partial \mathbf{u}}{\partial \mathbf{n}}\big{|}_{\Gamma_{\text{out}}}  = 0 \quad \text{Fully developed flow}
    \end{cases} \quad \begin{cases}
        p_{\mathrm{rgh}} \big{|}_{\Gamma_{\text{out}}}  = 10^5 \ \text{Pa}\\
        \frac{\partial p_\mathrm{rgh}}{\partial \mathbf{n}} \big{|}_{\Gamma_{\text{walls}} \cup \Gamma_{\text{cyl}} \cup \Gamma_{\text{in}}}= (\rho \mathbf{g} + \mathbf{J} \times \mathbf{B}) \cdot \mathbf{n}
    \end{cases}\\
    \begin{cases}
        T\big{|}_{\Gamma_{\text{in}}} = 600 \, \text{K} \\
        \frac{\partial T}{\partial \mathbf{n}}\big{|}_{\Gamma_{\text{walls}} \cup \Gamma_{\text{out}}} = \mathbf{0} \quad \text{Adiabatic walls}\\
        T\big{|}_{\Gamma_{\text{cyl}}} = 550 \, \text{K}
    \end{cases}
    \quad
    \begin{cases}
        \mathbf{B}\big{|}_{\Gamma_{\text{Side}} \cup \Gamma_{\text{cyl}}} = \mathbf{B}_{\text{ext}}  \quad \text{Insulator Side walls}\\
        \frac{\partial \mathbf{B}}{\partial \mathbf{n}}\big{|}_{\Gamma_{\text{Hart}} \cup \Gamma_{\text{out}}} = \mathbf{0} \quad \text{Conductive Hartmann walls}
    \end{cases}
\end{align}
completed with initial conditions:
\begin{equation}
    \begin{cases}
    \mathbf{u}(\mathbf{x}, 0) = |\mathbf{u}_0| \hat{x}  \\
    p_\mathrm{rgh}(\mathbf{x}, 0) = 10^5 \, \text{Pa} \\
    T(\mathbf{x}, 0) = 600 \, \text{K} \\
    \mathbf{B}(\mathbf{x}, 0) = \mathbf{B}_{\text{ext}}
    \end{cases} \qquad \forall \mathbf{x} \in \Omega.
\end{equation}
where $\mathbf{B}_{\text{ext}}$ is the external applied magnetic field, which provided its intensity $|\mathbf{B}_{\text{ext}}|$ and the inclination angle $\alpha$, described as:
\begin{equation}
    \mathbf{B}_{\text{ext}}=
     |\mathbf{B}_{\text{ext},y}| \sin (\alpha) +  |\mathbf{B}_{\text{ext},z}| \cos(\alpha)
\end{equation}

All the full order simulations were run for a time of 3 seconds, a sufficient time period to observe the complete, fully developed flow, with a sampling interval of 0.025 seconds, resulting in a total of 150 snapshots for each parametric scenario. The generic thermo-hydraulic field $\psi$ is rescaled according to the maximum and minimum values assumed across all specific parametric training configurations (each representing a unique combination of magnetic field intensity and inclination angle), using the equation\footnote{
In the following discussions, the tilde will be omitted for the sake of brevity, but all the fields must be considered rescaled according to Eq. \ref{rescalo}.}:
\begin{equation}
    \tilde{\psi}=\frac{\psi-\psi_{min}}{\psi_{max}-\psi_{min}}
    \label{rescalo}
\end{equation}

Regarding the measurements extracted from the high-fidelity simulations, only temperature measurement are used as input for the SHRED: the pressure and velocity field are estimated by performing indirect state estimation. Additionally, to simulate the electronic noise affecting the sampling procedure in real experimental facilities, the data are first rescaled according to Eq. \ref{rescalo} and subsequently polluted with a centered Gaussian noise $\varepsilon$ of standard deviation $\sigma$, such that $\varepsilon \sim \mathcal{N}(0,\sigma^2)$. In this work, a value of $\sigma=0.02$ has been adopted both for the mono- and double-parametric cases to introduce realistic measurement disturbances without obscuring the dominant features of the underlying signal.

During the training phase, SHRED learns the relationship between noisy temperature measurements and the latent dynamics of all thermo-hydraulic fields. In the test phase, once new noisy temperature measurements become available for a novel parametric scenario unseen during the offline stage, SHRED infers the full thermo-hydraulic state of the MHD system. Specifically, multiple SHRED neural networks are trained and tested using $4$ different temperature sensor random placements for each configuration.

To assess the accuracy of the ensemble reconstruction for a given field $\psi$, the mean relative error in the $L^2$ norm, $\epsilon_{L^2}$, is introduced as a metric of the average error as follows:
\begin{equation}
    \epsilon_{L^2}(\psi)= \left\langle\frac{\| \braket{\hat{\psi}}_K-\psi_{\text{FOM}}\|_{L^2}}{\| \psi_{\text{FOM}}\|_{L^2}} \right \rangle_{t \in \mathcal{T},\ \bm{\mu} \in \Xi^{\text{test}}_{\bm{\mu}}}
\end{equation}
where $\left\langle\cdot\right\rangle_{t \in \mathcal{T},\ \bm{\mu} \in \Xi^{\text{test}}_{\bm{\mu}}}$ represents the average across all time steps $t \in \mathcal{T}$ and all parametric scenarios $\bm{\mu} \in \Xi^{\text{test}}_{\bm{\mu}}$ in the test set.

Due to the shallow nature of the SHRED architecture, the training of each configuration requires approximately 3 minutes on a personal computer equipped with an Intel Core i7-13650HX processor for a lag value $L=30$, which is chosen because the relative time window describes a complete von Kármán shedding cycle, while the state estimation in the test phase is nearly instantaneous (less than 1 second), fulfilling the requirements for real-time applications. This represents a crucial improvement over computing full-order solutions, as each transient simulation requires more than six hours, depending on the number of processors across which the simulation is parallelized.

In the following subsections, the capability of the ensemble-SHRED approach in assessing both accuracy and trustworthiness is evaluated for mono-parametric cases, where either the inclination angle of the magnetic field varies at a constant magnitude, or the magnitude varies at a constant inclination angle, as well as for a double-parametric scenario in which both the intensity and inclination of the magnetic field vary simultaneously.

\subsection{Single Parameter Analysis}

While SHRED has already been proven to accurately capture the non-linear relationship between temperature measurements and the reduced dynamics of MHD systems for mono-parametric cases in a simpler geometry \cite{loverso2025reduced}, this section demonstrates both the accuracy and the trustworthiness of the state estimation, for a significantly more challenging MHD benchmark case (both from an engineering perspective and due to the intrinsic difficulty arising from the interaction between von Kármán vortices and side layers). 

\subsubsection{Single Parameter Analysis: Varying the Magnetic Field Inclination}

First, the reconstruction accuracy for unseen inclination angles is assessed, along with its improvement as a function of the training set size. According to \cite{unknown1,TheDEMO}, the total magnetic field generated by the magnets, in a Tokamak system, is predominantly toroidal with a small poloidal component. Therefore, the inclination angle in almost all operative conditions ranges between 5 and 30 degrees. Thus, the system has been simulated for 17 different angles spanning the given range, keeping the same magnetic field magnitude $\mathbf{B}_\mathrm{ext} = 0.05 $T. This value is sufficiently low to avoid complete laminarization of the flow.

Regarding the angle value, let $\Xi^\alpha$ be the set of simulated values:
\begin{align*}
      \Xi^{\alpha}= \{ &\text{5°, 6.4°, 7.8° , 9.2°, 9.5°, 10.6°, 13.3°, 14.7°, 17.5°, 18.9°, 20.3°, 21.7°, 23.1°, 24.4° , 27.2°, 28.6°, 30°} \}.
\end{align*}

As will be shown in the following, 17 angles constitute a very large sample to represent the reduced dynamics for such a small range. Regardless, such a large training dataset is useful to prove the data efficiency of SHRED with respect to the size of the training set, and hence to study the saturation trend of the mean relative error as a function of the number of angles in the training set. The validation set $\Xi^\alpha_{\text{valid}}$ test set $\Xi^\alpha_{\text{test}}$ have been chosen to be representative of the entire range, and hence:
$$ \Xi_\alpha^{\text{test}}=\{\text{7.8°, 18.9°, 27.2°} \} \qquad  \Xi_\alpha^{\text{valid}}= \{ \text{6.4°, 14.7°, 28.6°}\}$$

Seven different training sets have been used, each containing a different number of training angles, to train seven separate neural networks. 
\begin{figure}[htbp]
    \centering
    \includegraphics[width=0.75\linewidth]{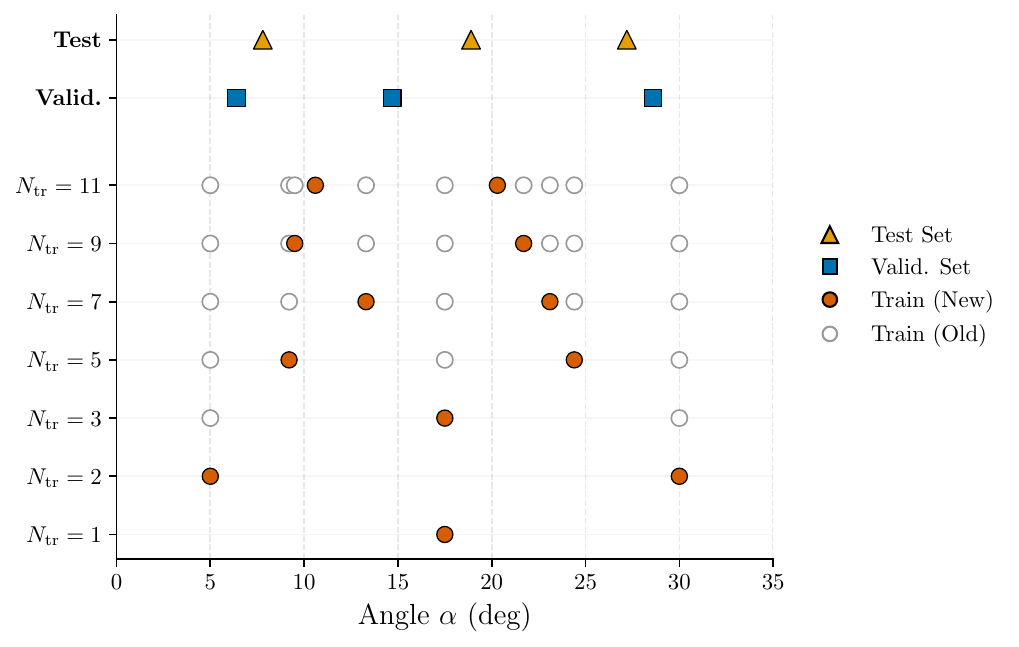}
    \caption{Train, test and validation splitting. Training angles, increasing $N_{\text{tr}}$, are added to the previous set of training angles, with the exception of the first two training set with only one and two angles.}
    \label{dataa}
\end{figure}
In Figure \ref{dataa}, the increasing number of training angles $N_{\text{tr}}$ is shown together with the test set and train set. The first two training sets, consisting of only one angle, 17.5°, in the middle of the range, and of only two angles, 5° and 30°, coinciding with the extremes of the range, are used to assess the capabilities of SHRED when exploiting a very small number of training cases and understanding its generalization capabilities in the low-data limit. In particular the first case studies whether SHRED, once it has seen the behaviour at the average angle of the total set, is able to infer the dynamics near the extremes, while the second case assesses whether SHRED is capable of inferring the average dynamics once it has seen the dynamics at the extremes of the range.

\begin{figure}[hbpt]
    \centering
    \includegraphics[width=1\linewidth]{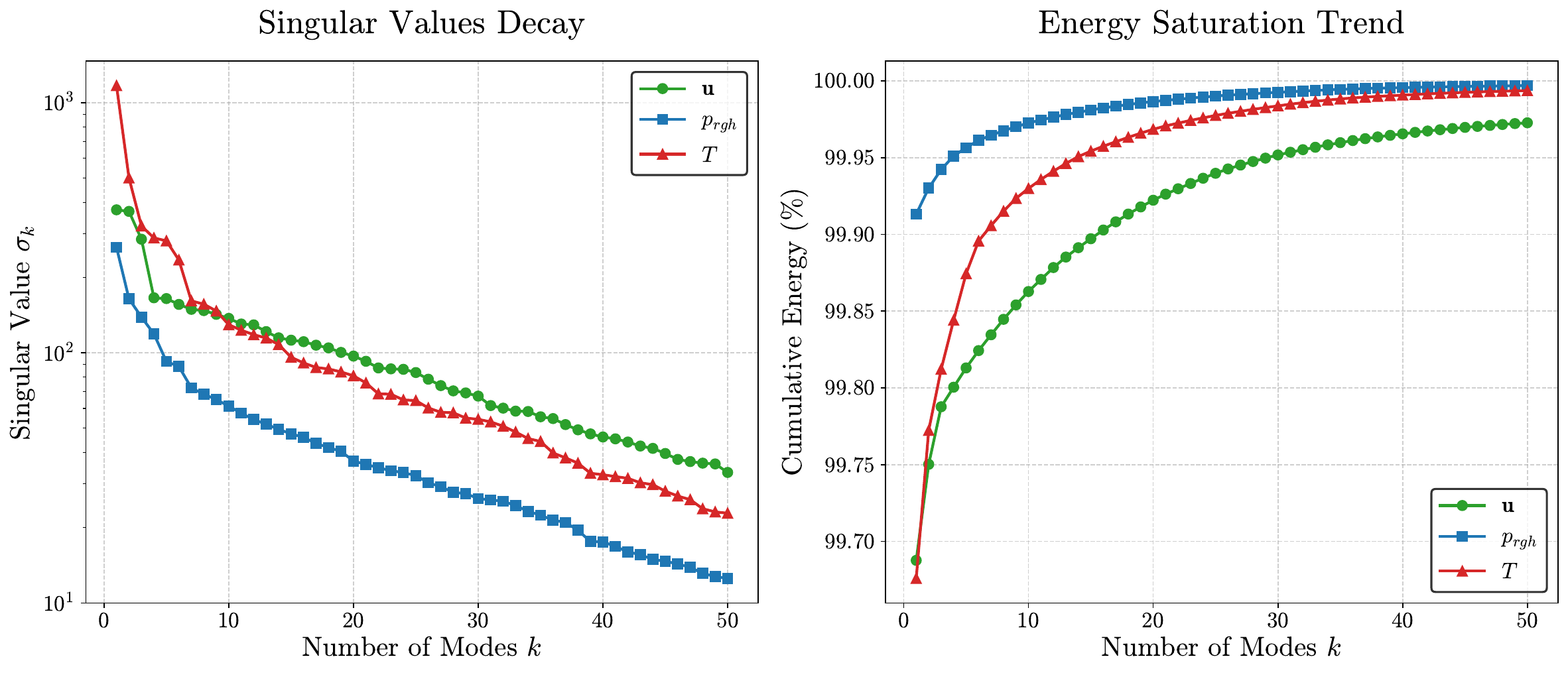}
    \caption{Principal component analysis on training data set for $N_{tr}=11$.}
    \label{PCA_fine}
\end{figure}

The first preliminary analysis involves assessing whether the solution manifold for this case can be approximated with a linear subspace, spanned by the PCA modes. The decay of the singular values, according to PCA, is shown in Fig.~\ref{PCA_fine}. It is evident that even few modes are capable of accurately approximating the solution manifold, spanning the reduced basis. In particular, since the average snapshot has been removed, the PCA modes capture the dynamics of the fluctuations, and thus in order to compute the total cumulative energy encoded by the $k$ modes, the energy of the averaged snapshot has to be added. As can be seen, even a single PCA mode, when considering the average field, encodes more than 99.6\% of the total information for each field. Nevertheless, the remaining 0.4\% encodes the more complex dynamics, such as small turbulent structures; as expected, the velocity field shows the slowest saturation trend, since it encodes more complex dynamics than the pressure and temperature fields. Hence, in order to capture them within a dimensionality reduction context, a compromise truncation rank of 20, $r=20$, has been chosen for each field.

As aforementioned, the SHRED state estimation is \textit{agnostic} to the position of the sensors placed in the domain, meaning that the performance is not significantly affected by different sensor locations (provided that the output measurements contain enough temporal statistics). Moreover, the ensemble approach allows the reconstruction to be smoothed with respect to disturbances, which may be dictated by sensor noise or by placements in regions where the dynamics are not particularly relevant, by averaging the different state estimations obtained from diverse sensor networks \cite{riva2024robuststateestimationpartial}. 

To prove sensor agnosticism, candidate sensor locations are either randomly distributed or selected by the DEIM algorithm \cite{quarteroni_reduced_2015}. The DEIM algorithm selects sensor placements that encode the maximum amount of information regarding the spatial patterns. Consequently, placing sensors at the selected DEIM positions maximizes the information content extracted from the temperature field. It is thus expected that DEIM-SHRED could yield a reconstruction accuracy superior to the one obtained with randomly placed sensors, particularly when only a few sensors (e.g., only four) are available. SHRED agnosticism is considered demonstrated if the mean relative errors computed using either randomly chosen sensors or those selected by DEIM are nearly identical.

Even when exploiting the DEIM sensor placement strategy, the ensemble approach is still a smart choice, since it provides, through the standard deviation field, an estimate of the reconstruction error \cite{riva2024robuststateestimationpartial}. Regions that are more difficult to reconstruct and hence more sensitive to sensor positions can be identified as regions that are harder to estimate and therefore likely to exhibit higher errors. Accordingly, in the following analysis, once the best 10 sensor positions have been computed by DEIM, 5 configurations, each sampling four sensors, are implemented for the ensemble.

\begin{figure}[htbp]
    \centering
    \includegraphics[width=1\linewidth]{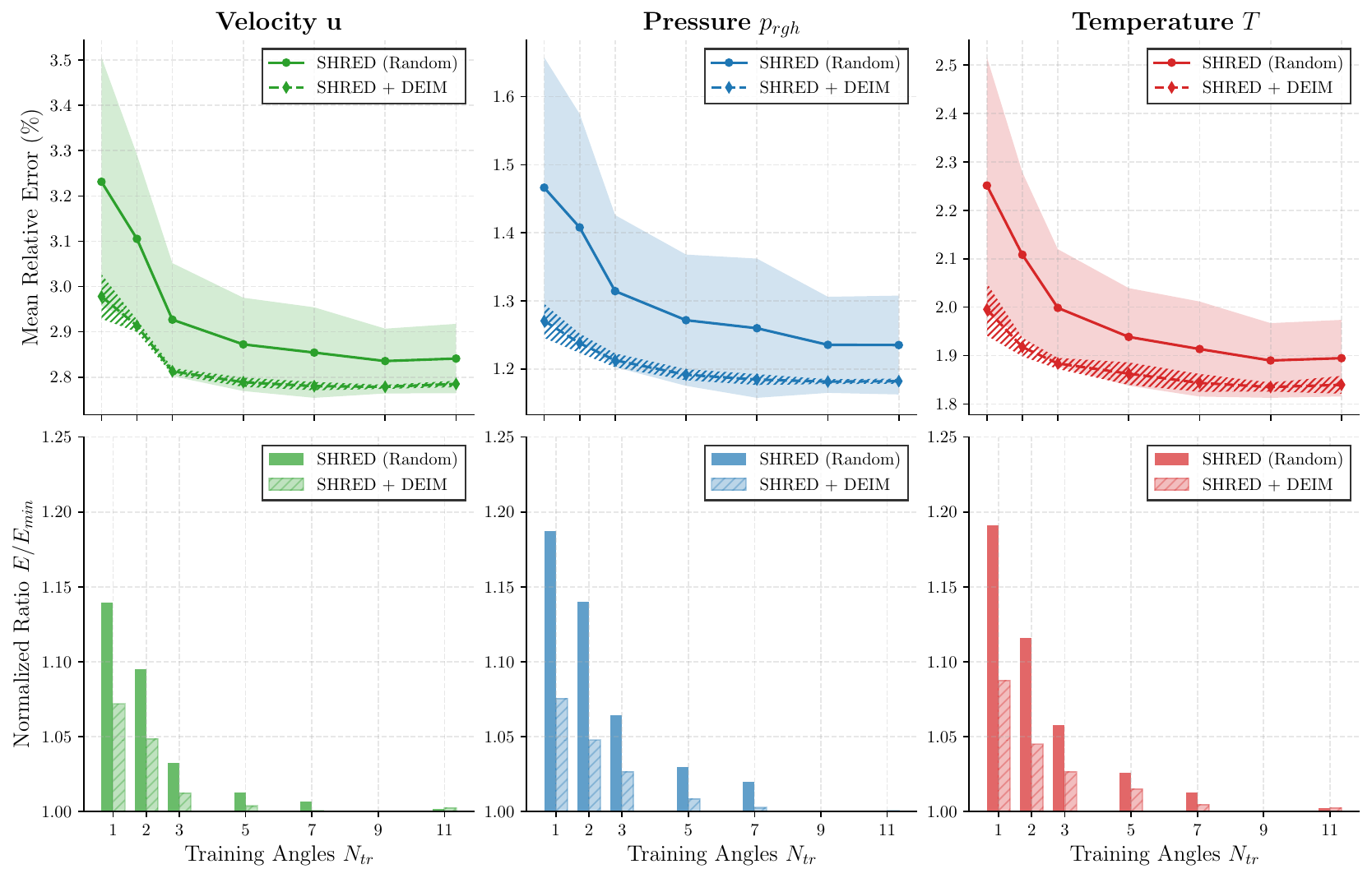}
    \caption{On top, the mean relative errors for the velocity, pressure, and temperature fields are shown as a function of the number of training angles, both for sensor positions randomly chosen, according to the agnosticism of SHRED, and for the optimal sensor placement determined by the DEIM algorithm. Below, the saturation behavior of the $mre$ as the number of angles increases is represented, expressing the data efficiency of SHRED and its generalization capability.}
    \label{shreddino}
\end{figure}

In Figure \ref{shreddino}, the overall comparison of the mean relative error for all fields across the dimension of the training set for both the standard SHRED and DEIM-SHRED is presented. Regardless of the sensor placement strategy, the reconstruction accuracy is evident, as the mean relative error is consistently lower than 3.3\%. This means that even with limited data, SHRED is able to infer the overall dynamics, and hence it is not merely memorizing but is actually understanding the correlation between real measurements and the latent space spanned by the PCA. This proves the SHRED data efficiency, thereby proving its generalization capability. From an engineering perspective, the possibility of investigating novel scenarios given only a few high-fidelity simulations is a massive advantage, as the number of computationally expensive simulations is significantly reduced, while still guaranteeing sufficient accuracy to ensure the precision and reliability of the state estimation.

Nevertheless, the combination of DEIM and SHRED results in a mean relative error that is consistently lower than that of the standard SHRED: not only is DEIM-SHRED more accurate, but its mean relative errors also saturate earlier as a function of the number of training angles; specifically, even just three training angles, for the DEIM-SHRED, are sufficient to completely saturate the mean relative error, as it can be seen in the plot of the ratio of the mean relative error relative to the lowest mean relative error. Regarding standard SHRED, the saturation point is shifted toward a higher number of training angles. Nevertheless, the actual difference in mean relative error between SHRED and DEIM-SHRED is always less than 0.25\%, demonstrating the agnosticism of SHRED with respect to sensor positions. Moreover, alongside the mean relative error of the average reconstruction obtained using the ensemble approach, the lowest and highest errors corresponding to the best and worst sensor configurations are also shown. In particular, as anticipated, since the candidate sensors for DEIM-SHRED are optimal, outliers are avoided, while for the standard SHRED certain sensor arrangements in the domain can result in reconstruction errors that are larger or smaller than typical. As can be clearly seen, the mean relative error of the ensemble reconstruction is almost always approximately midway between the best-case and worst-case scenarios.

\begin{figure}[htbp]
    \centering

    \subfloat[]{
        \includegraphics[width=0.8\textwidth]{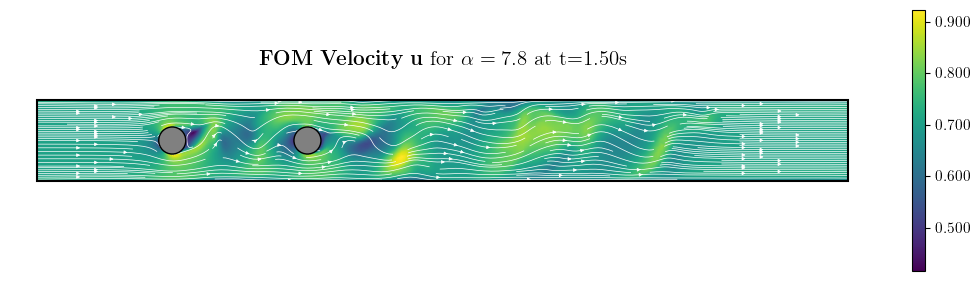}
    }

    \vspace{0.5cm}

    \subfloat[]{
        \includegraphics[width=0.45\textwidth]{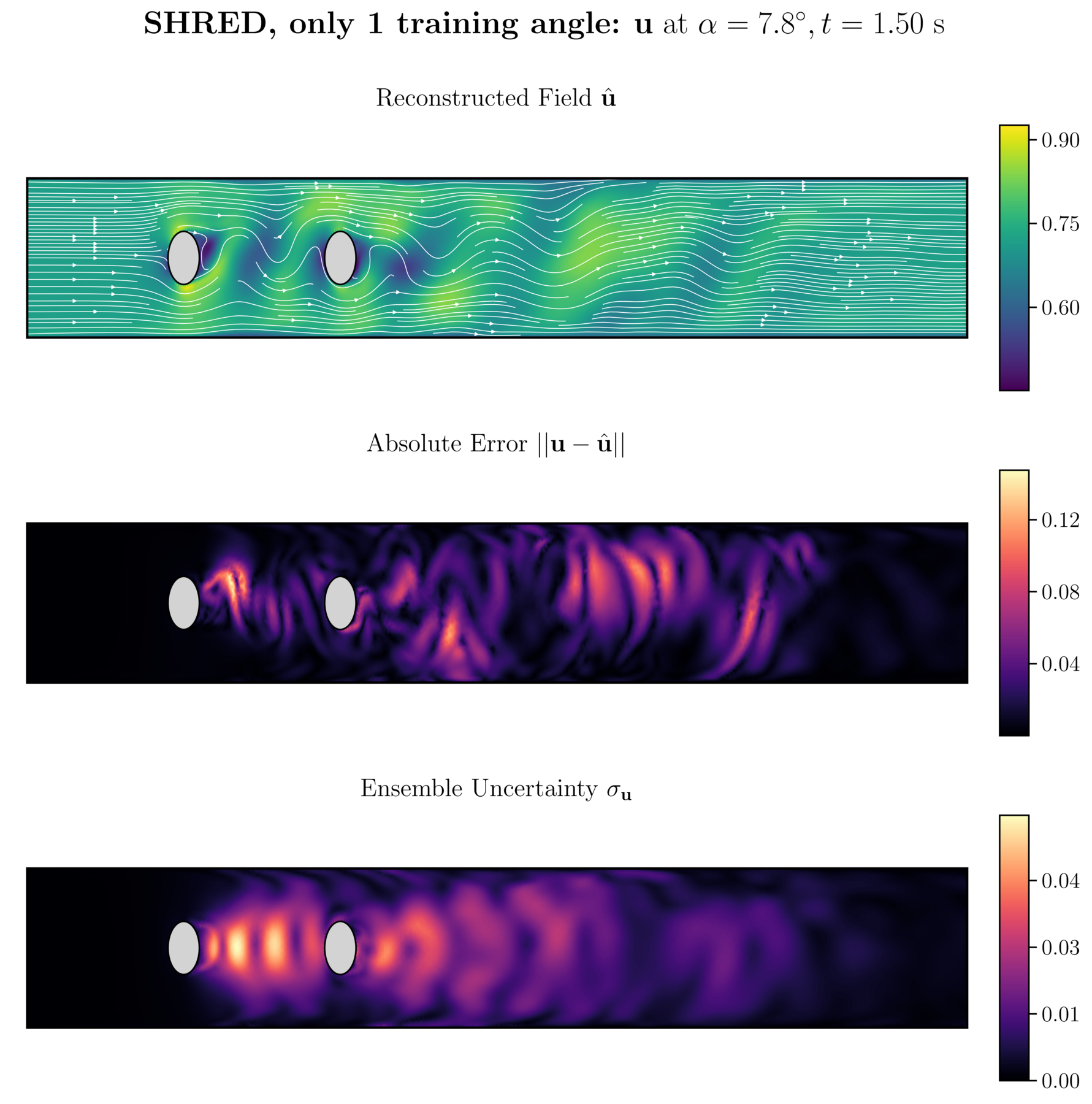}
    }
    \hfill
    \subfloat[]{
        \includegraphics[width=0.45\textwidth]{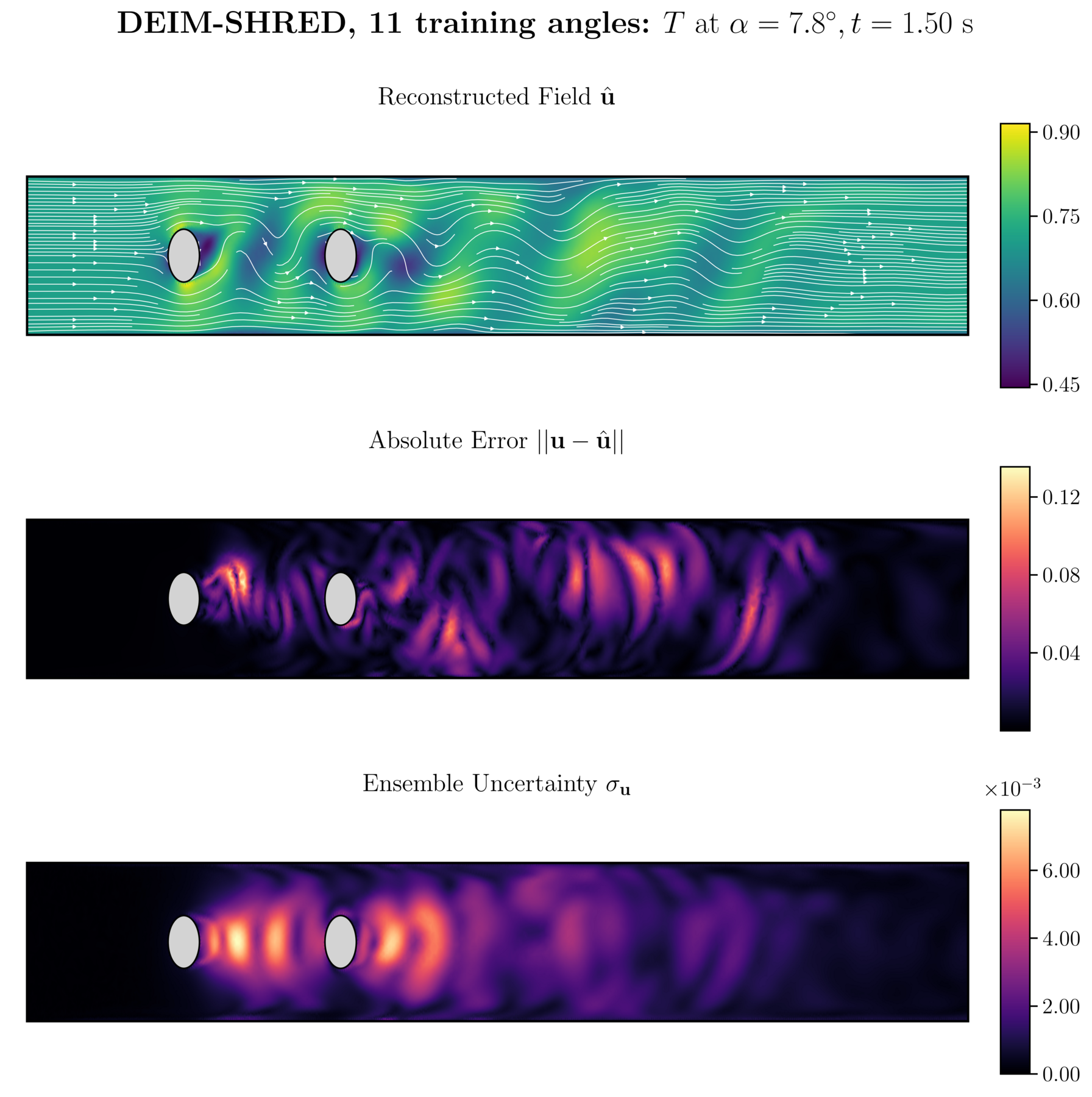}
    }

    \caption{Comparison of the normalized velocity field between DEIM-SHRED with 11 training angles and standard SHRED with only one training angle $\alpha_{train}=$17.5°.}
    \label{fig:piramide1}
\end{figure}

\begin{figure}[htbp]
    \centering

    \subfloat[]{
        \includegraphics[width=0.8\textwidth]{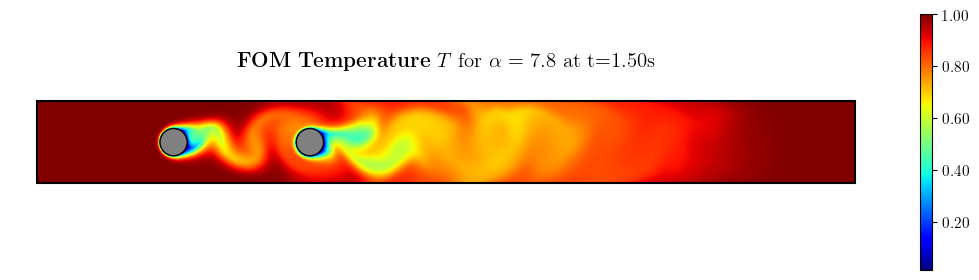}
    }

    \vspace{0.5cm}

    \subfloat[]{
        \includegraphics[width=0.45\textwidth]{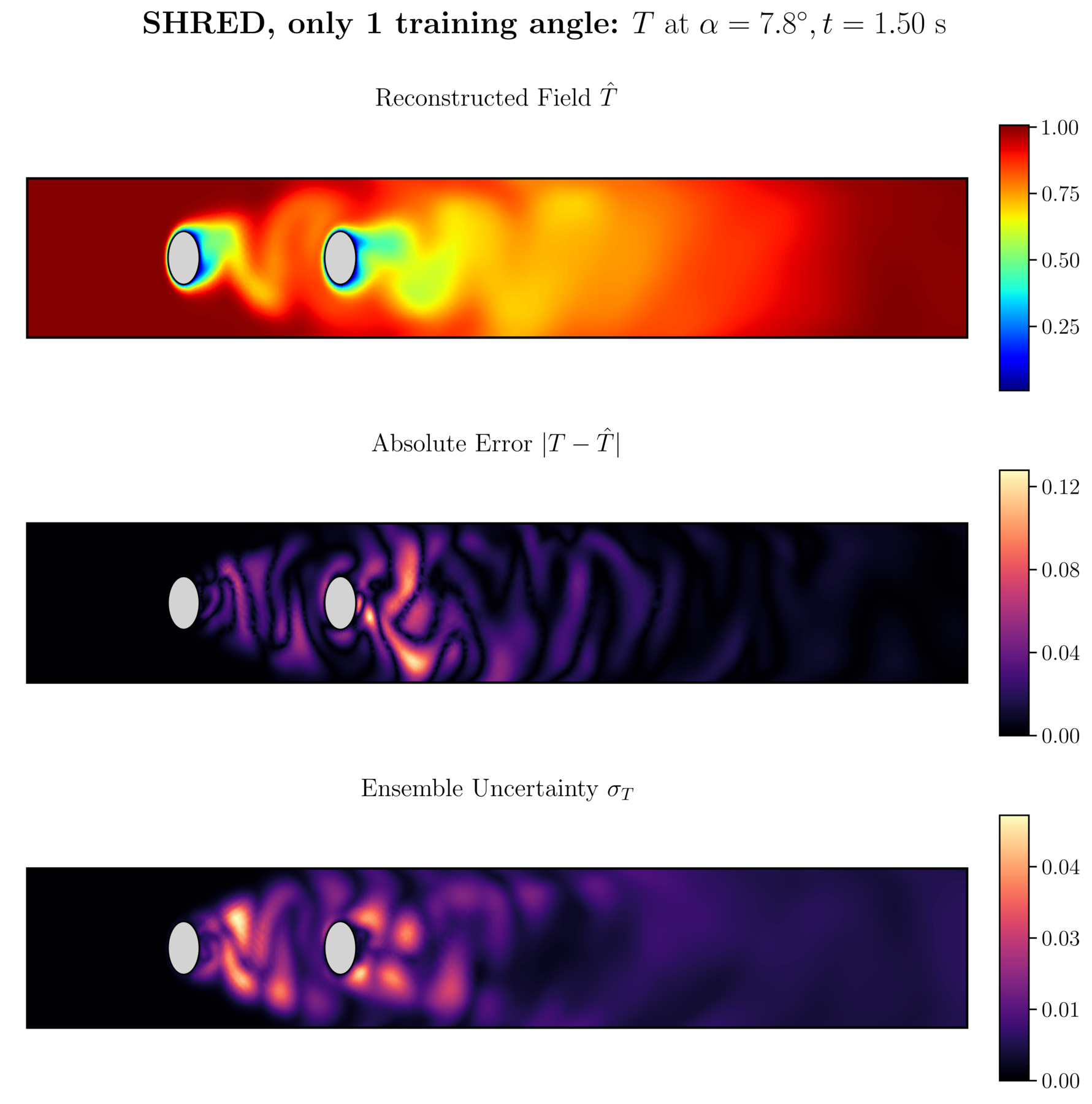}
    }
    \hfill
    \subfloat[]{
        \includegraphics[width=0.45\textwidth]{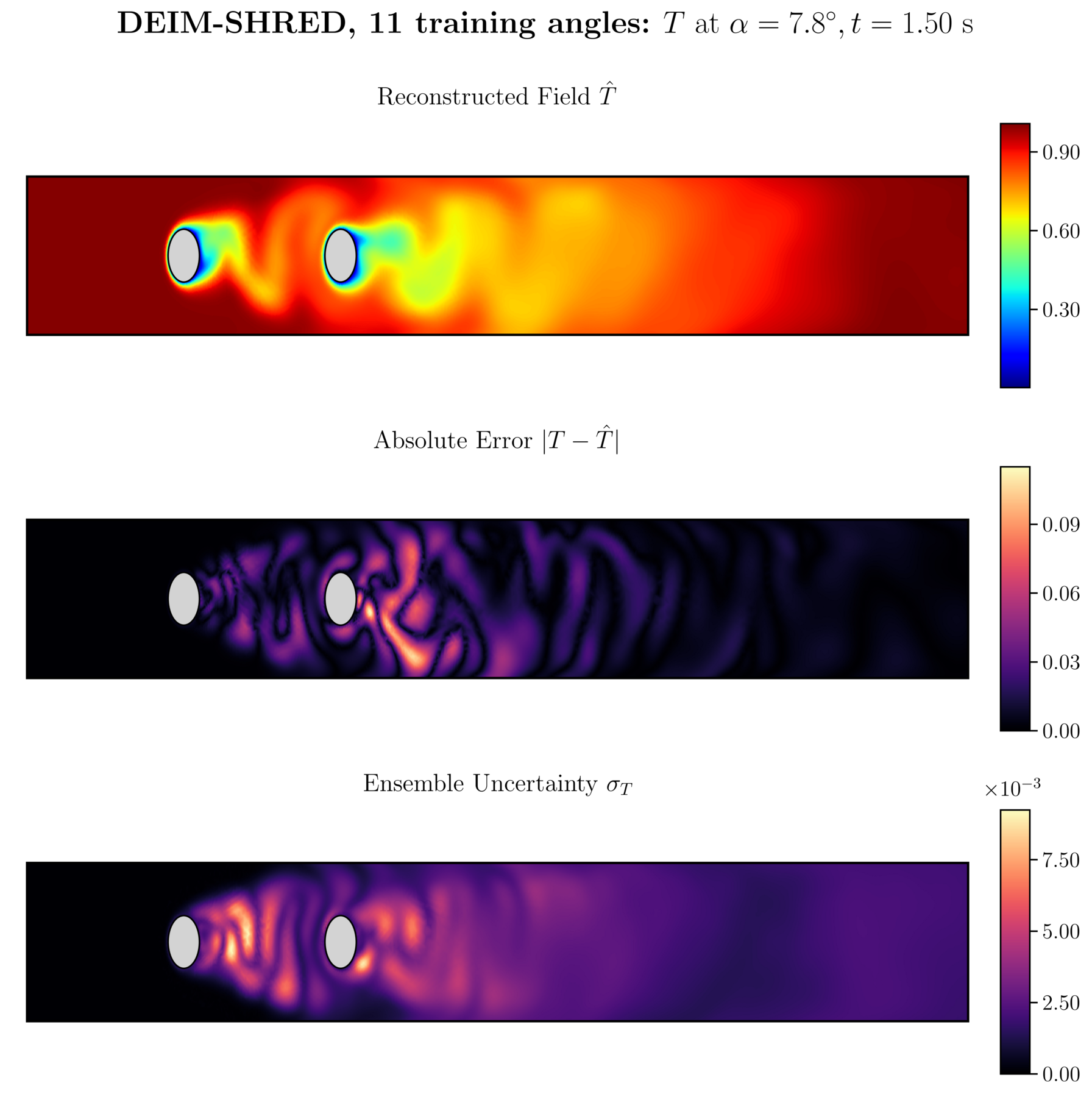}
    }

    \caption{Comparison of the normalized temperature field between DEIM-SHRED with 11 training angles and standard SHRED with only one training angle $\alpha_{train}=$17.5°.}
    \label{fig:piramide}
\end{figure}

In conclusion, to effectively visualize the state estimation capability of SHRED when using randomly placed sensors and only one training angle (17.5°), a comparison with the state estimation obtained by DEIM-SHRED using 11 training angles (the best-case scenario) is shown in Figure \ref{fig:piramide1} (for the velocity field) and Figure \ref{fig:piramide} (for the temperature field). From a qualitative perspective, no remarkable differences between the two state estimations are observed, thereby substantiating sensor agnosticism and generalization capability for SHRED.

In particular, the test snapshot is at an angle value that is not close to the training angle, otherwise the comparison could be questioned, as the unseen data is very close to the training data. In the same figures, the residual distributions and the standard deviation distributions are also shown. As previously stated, the standard deviation field acts as an estimator of the regions in the domain where the highest errors are more likely to be found. In particular, downstream of the cylinders, in the presence of von Kármán vortices \cite{davidson2015turbulence}, the SHRED predictions are less accurate, as predicted by the ensemble uncertainty. In order to represent the solutions in the bulk of the flow, the longitudinal mid-plane cut has been considered for plotting the fields. 

Regarding the pressure field, as illustrated in Fig. \ref{PCA_fine}, since even a single PCA mode captures more than 99.9\% of the total information of the pressure dynamics, a sufficiently accurate approximation is to consider just the average static pressure field as a trustworthy representation. This field ultimately exhibits a quasi-static behavior rather than complex dynamics. Consequently, presenting specific pressure plots would not provide additional insights. Nevertheless, as shown in Figure \ref{shreddino}, SHRED was still employed to estimate the full pressure field, rather than limiting the evaluation to the simple representation provided by the average field. From a physical perspective, the quasi-static behavior of the pressure field can be attributed to the boundary conditions of the FOM simulations in relation to the low flow velocity. Specifically, the low velocity magnitude generates dynamic pressure fluctuations that are orders of magnitude smaller than the fixed pressure applied at the outlet; thereby, the overall pressure variance is negligible compared to the static background pressure. Moreover, by applying the divergence operator to the Navier-Stokes momentum equation, the following \textit{Poisson} equation for the pressure is obtained:
\begin{equation}
    \nabla^2 p_{rgh} = - \rho \nabla \cdot (\mathbf{u} \cdot \nabla \mathbf{u}) + \nabla \cdot (\mathbf{J} \times \mathbf{B}) - \nabla \cdot [ (\mathbf{g} \cdot \mathbf{h})\nabla\rho ]
\end{equation}

As the magnetic field increases, the Joule damping effect intensifies, suppressing turbulent eddies and thereby rendering the convective term $\rho \nabla \cdot (\mathbf{u} \cdot \nabla \mathbf{u})$ negligible. Furthermore, due to the low magnetic Reynolds number typical of MHD Tokamak systems, the self-induced magnetic field is negligible compared to the externally imposed one, which remains constant \cite{buhler2007liquid, wesson2011tokamaks}. Since liquid metals are characterized by a Prandtl number much lower than one, the temperature field is extremely smooth, implying that the buoyancy source term is also negligible ($\nabla \cdot [ (\mathbf{g} \cdot \mathbf{h})\nabla\rho ] \propto \nabla^2 T \approx 0$). Finally, because the externally applied magnetic field and the principal velocity direction are perpendicular and constant in the core flow, the Lorentz force $\mathbf{J} \times \mathbf{B}$ is uniform, and hence, its divergence vanishes. The suppression of all source terms in the Poisson equation implies that the pressure field lacks local spatial curvature. Combined with the applied boundary conditions, this dictates an overall quasi-static pressure field. Moreover, as the intensity of the magnetic field increases, pressure homogenization becomes more pronounced.

\subsubsection{Single Parameter Analysis: Varying the Magnetic Field Intensity}

This section assesses the ensemble-SHRED capability to infer the flow dynamics for unseen magnetic field intensities, examining its performance improvement as a function of training set size, as a precursor to the two-parameter analysis. Moreover, particular attention is devoted to the interpretation of the results, and hence the mean relative error, individually for each value of the magnetic field intensity in the test set. In particular, 21 different high fidelity simulations at different values of the magnetic field intensity have been simulated for the same fixed inclination angle equal to 5°, spanning different regimes. The inclination angle has been chosen with reference to the Hunt flow, in which the magnetic field is transverse to the flow and parallel to the upper and lower walls. In particular, the present study investigates whether SHRED is capable of predicting the formation of symmetric side layers.

\begin{figure}[hbpt]
    \centering
    \includegraphics[width=0.6\linewidth]{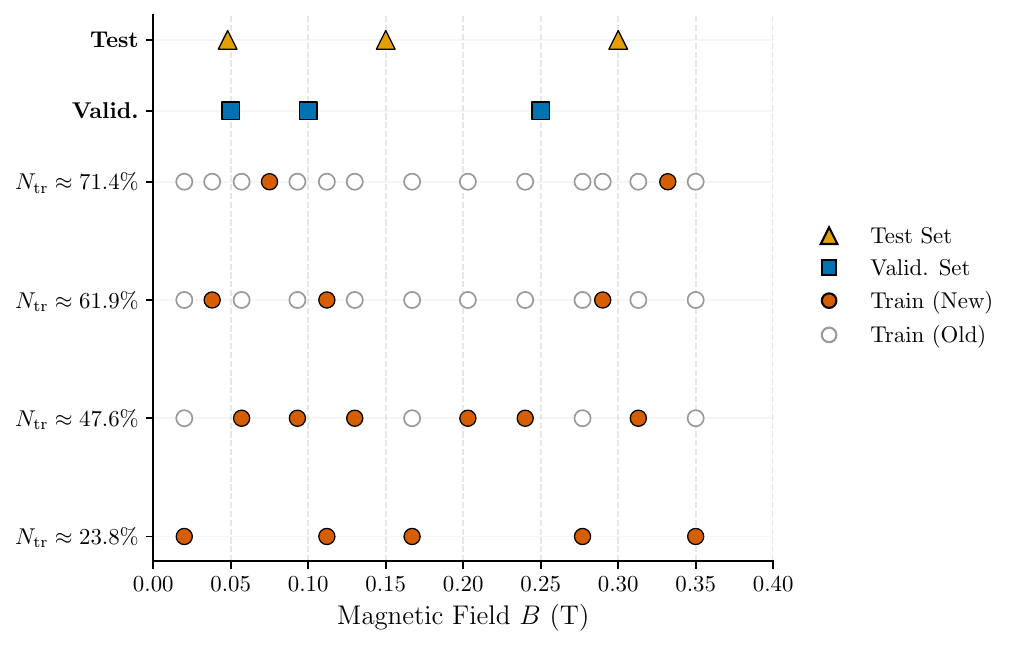}
    \caption{Train, test and validation splitting for the magnetic field intensity. Increasing $N_{\text{tr}}$, new training cases are added to the previous set of training magnetic fields.}
    \label{training_set}
\end{figure}

Four different training sets have been chosen in order to investigate SHRED performance on training sets of sizes corresponding approximately to 25\%, 50\%, 60\%, and 70\% of the total simulated magnetic field cases. These sets were constructed hierarchically, such that larger training sets contain the magnetic field values from the smaller ones, as summarized in Figure \ref{training_set}. 


Regarding the intrinsic difficulty of this reconstruction task, it can be argued that, while the dependence of the solution on the inclination angle is relatively weak, the dependence on the magnetic field intensity is significantly stronger. In this case, the dynamics are strongly affected by the magnetic braking effect induced by a high-intensity magnetic field \cite{article}. In particular, varying the inclination angle while keeping the magnitude of the magnetic field constant results in a constant amount of magnetic energy introduced into the system: changes in the dynamics are simply due to a different spatial distribution of this energy, and it is expected that mainly the spatial patterns will differ, while the temporal dynamics are not strongly affected. On the contrary, keeping the inclination angle constant and varying the intensity changes the energy introduced into the system, and hence not only is the dynamics perturbed in its form, but new flow regimes are established, characterized by completely different spatial and temporal patterns.

\begin{figure}[hbpt]
    \centering
    \includegraphics[width=1\linewidth]{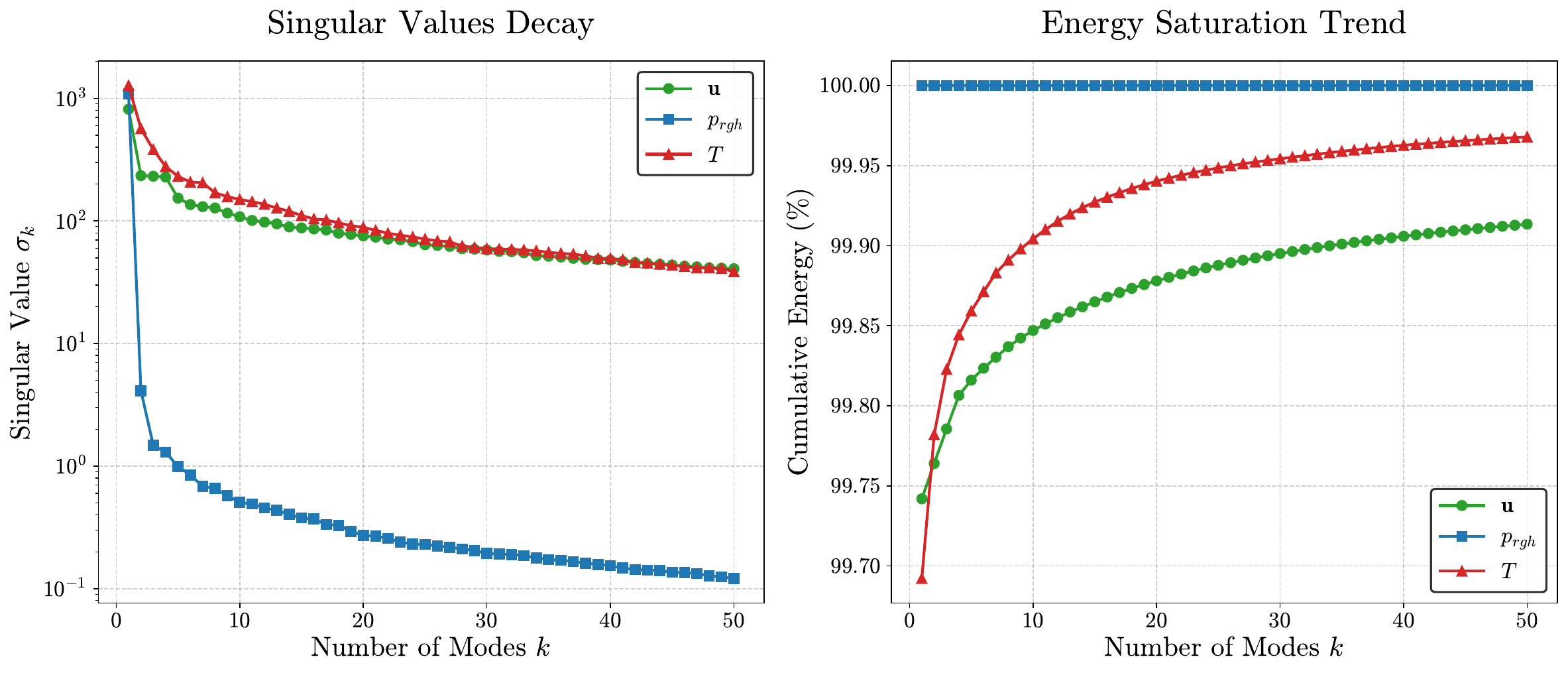}
    \caption{Principal component analysis on training data set for $N_{tr}=15$.}
    \label{mostra_fotografica}
\end{figure}

\begin{figure}[htbp]
    \centering
    \includegraphics[width=1\linewidth]{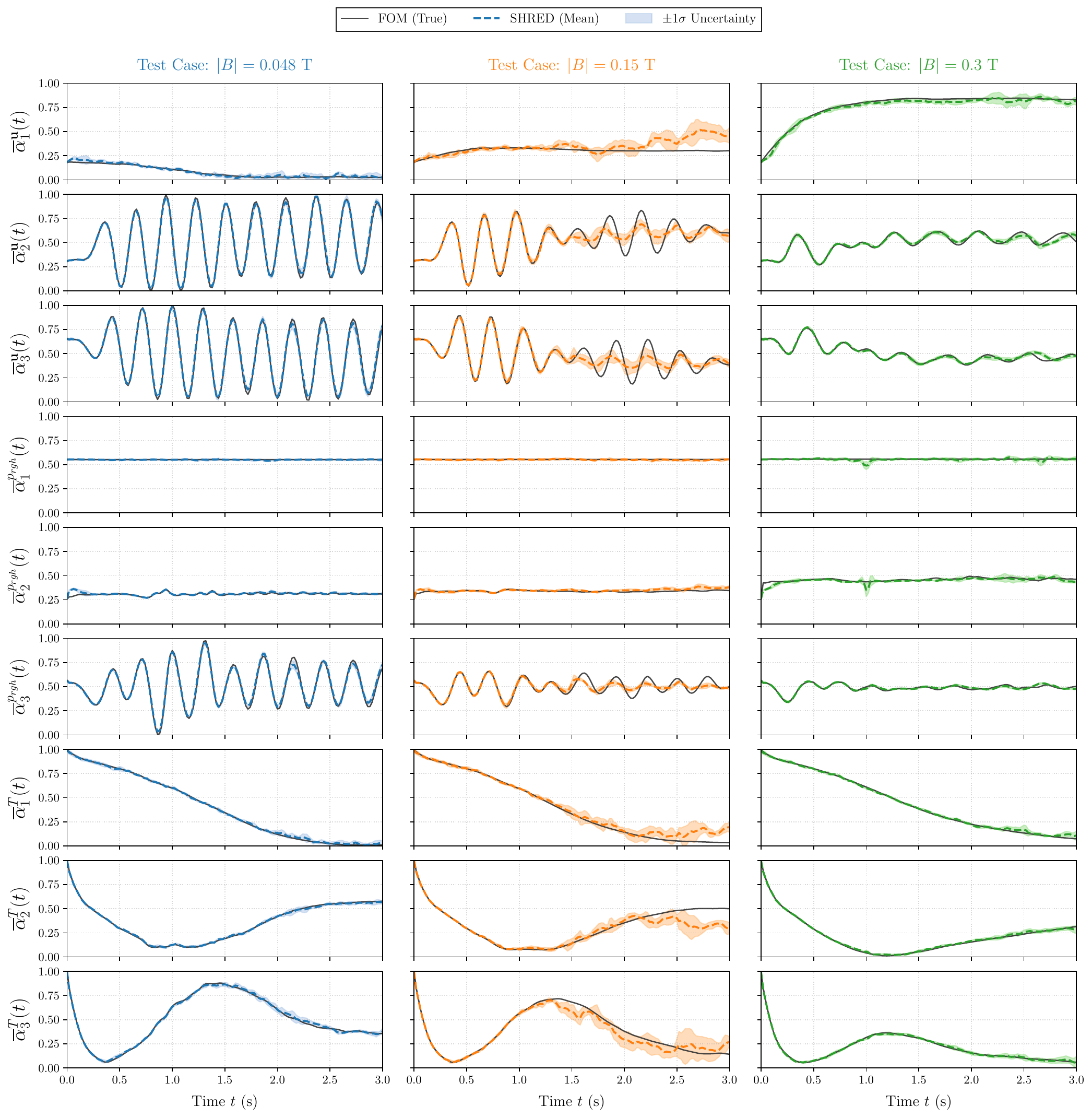}
    \caption{Temporal evolution of the first three PCA coefficients, as predicted by the SHRED, for the different fields and different values of magnetic field test. As expected for the pressure field, the first two PCA coefficients are completely flat, highlighting the absence of temporal dynamics, with the exception of low-importance dynamics appearing in subsequent modes, according to the hierarchy of the PCA modes.}
    \label{mega_plot}
\end{figure}

The higher parametric complexity of this second scenario is confirmed by the decay of the PCA singular values and by the cumulative energy trend, as shown in Fig.~\ref{mostra_fotografica}. 
As aforementioned, since the training set contains specific cases with an intense external magnetic field applied ($B_{\text{ext}} > 0.3 \ \text{T}$), the pressure field tends to be more spatially uniform and nearly constant in time compared to the first parametric case, in which the inclination angle varied at a low magnetic field magnitude ($B_{\text{ext}}=0.05 \ \text{T}$). This behavior is demonstrated by the fact that a single PCA mode captures nearly $100\%$ of the total information. In this context, the average pressure field across all test cases can be computed and assumed, with a high level of confidence, to be a sufficiently accurate representation of the test cases. This effect is shown, together with the prediction of the latent dynamics of the velocity and temperature fields, in Figure \ref{mega_plot}. Together with the prediction of the PCA coefficients, the ensemble-SHRED also provides an estimate of their uncertainty, according to the latent dynamics estimated from different sensor positioning. The overall narrowness of the standard deviation shadows means that different SHRED networks, trained by exploiting temperature measurements coming from different sensors, provide almost identical predictions, proving once more the agnosticism of SHRED.
From the PCA temporal evolutions, it can be expected that the most difficult test case to reconstruct is the one parametrized by $B_{\text{ext}} = 0.15\ \text{T}$. Indeed, with reference to Figure ~\ref{training_set}, no additional training cases have been added close to it. On the contrary, for the other two test cases, as the number of training magnetic field values increases, the distance from the test set is unavoidably reduced. Accordingly, and in addition to the fact that it corresponds to the average magnetic field of the total magnetic set $\Xi^{B_{\text{ext}}}$, the test case with $B_{\text{ext}} = 0.15\ \text{T}$ is considered an indicator of the overall SHRED performance, in particular regarding the saturation of information as the training set size increases and its capability to estimate the average behavior of the system for the average magnetic field. Nevertheless, the other two test cases are also relevant, as they represent completely different regimes of the liquid metal flow: for $B_{\text{ext}} = 0.048$ T, the magnetic influence is not sufficient to dominate turbulence and, hence, a chaotic flow is established; on the contrary, for $B_{\text{ext}} = 0.3$ T, as theoretically expected for a Hunt flow under a non negligible external magnetic field, laminarization is observed together with the presence of two symmetric side layers which, due to their high velocity, carry the vast majority of the mass flux.

\begin{figure}[htbp]
    \centering
    \includegraphics[width=1\linewidth]{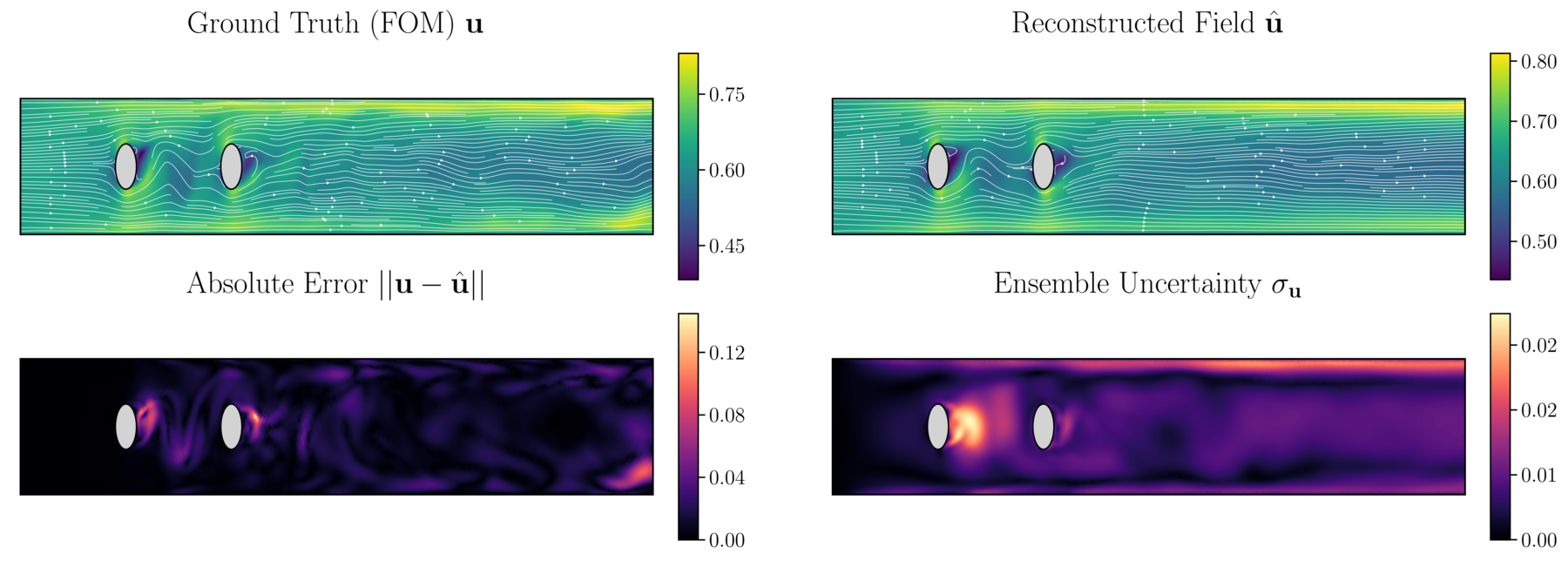}
    \caption{SHRED normalized reconstruction of the velocity field for $B_{\text{ext}} = 0.3$ T at $t = 2.5$ s, together with the absolute error and ensemble uncertainty.}
    \label{bella}
\end{figure}

In Figure ~\ref{bella}, the capability of SHRED to accurately reconstruct this characteristic MHD phenomenon is shown. The neural network effectively captures the formation of two symmetric side layers. Moreover, the ensemble uncertainty is higher in the regions where the suppression of von Kármán vortices behind the cylinders is acting due to the high magnetic field, as these are the regions that are more difficult to reconstruct. Accordingly, when computing the absolute error, the highest values are found exactly where the ensemble uncertainty is higher, proving the role of ensemble uncertainty as an estimator of the regions in which the reconstruction presents higher errors.

\begin{figure}[htbp]
    \centering
    \includegraphics[width=1\linewidth]{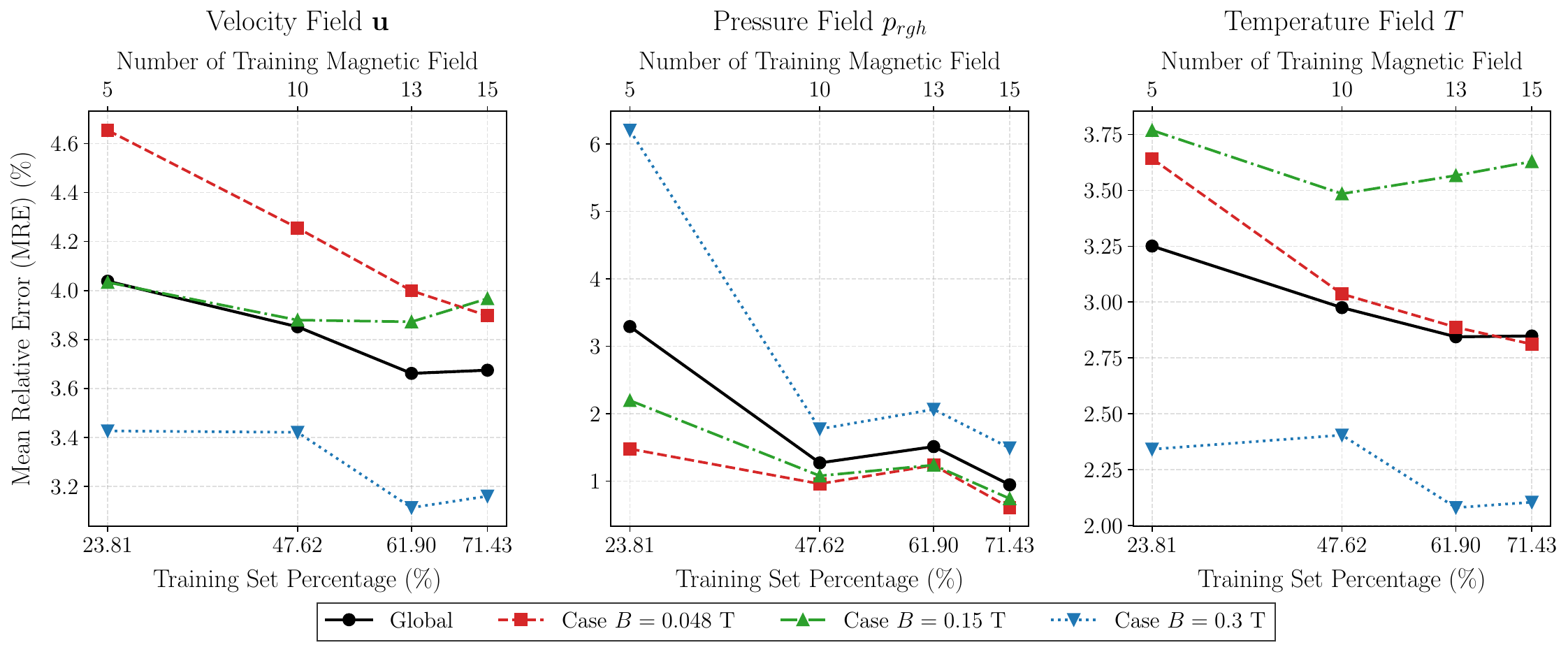}
    \caption{Mean relative errors for each field and for each value of the test magnetic field intensity, as a function of the relative training set size.}
    \label{mega_plottino}
\end{figure}

In Fig.~\ref{mega_plottino}, the mean relative errors for each thermo-hydraulic field and for each value of the magnetic field intensity, as a function of the relative training set size, are shown. As can be expected, as the number of magnetic field values in the training set increases, the mean relative error globally decreases for each field. Regarding its value for the velocity field, it can be noted that as the value of the test magnetic field increases, the error committed decreases. As expected, a higher magnetic field laminarises the flow, suppressing turbulence and hence making it easier for the neural network to estimate the state. On the contrary, for low values of the magnetic field, turbulence may be interpreted by SHRED as noise, and hence the performance slightly worsens. Interestingly, regarding the pressure field the opposite behavior is observed. In particular, as the external magnetic field becomes more intense, the Joule damping, proportional to the square of its intensity through the Lorentz force $\mathbf{f}_L \propto B_{\text{ext}}^2$, dictates a more intense magnetic brake, which has to be counterbalanced by a higher pressure gradient in order to push the liquid along its main flow direction. The neural network struggles slightly more to interpolate fields across such drastically scaled magnitudes, which is reflected in higher relative errors.

As for the temperature field, it can be noticed that for the case $B_{\text{ext}} = 0.3$ T the mean relative error curve is lower than for the other two test cases. From a physical perspective, for intense magnetic fields, due to the laminarization effect, the temperature field is strongly coupled with the velocity field according to the advection term $\nabla \cdot (\rho \mathbf{u} T)$, as heat is transported in an orderly fashion by the steady velocity streamlines \cite{bird2006transport}. This strong coupling between the temperature and velocity fields at high magnetic field intensities is clearly reflected in the state estimation performance. In particular, the decreasing trends of the mean relative errors for $B_{\text{ext}} = 0.3 \ \text{T}$ for both velocity and temperature are strikingly similar. Accordingly, this suggests that SHRED effectively can capture and preserve the underlying physical correlations between different thermo-hydraulic fields learned during the training phase. Moreover, regarding $B_{\text{ext}} = 0.15$ T, and hence the test case with the same distance from the closest training set even as the training set size increases, the mean relative error curve is always close to those of the other two test cases. As the mean relative error is systematically lower than 5\%, SHRED captures the overall dynamics remarkably well by learning the temporal evolution of spatial patterns, rather than memorizing information and simply fitting the data.

In conclusion, the generalization capability of SHRED is further proven, as the mean relative error for all the fields when considering training set percentages of 20\% and 70\% remains almost the same. The sole exception occurs in the pressure field at the highest magnetic field intensity ($B_{\text{ext}}=0.3\ \text{T}$), where the mean relative error nevertheless does not exceed 6.5\%.
\subsection{Double Parameter Analysis: Varying the Magnetic Field Intensity and Inclination}

\begin{figure}[hbpt]
    \centering
    \includegraphics[width=0.7\linewidth]{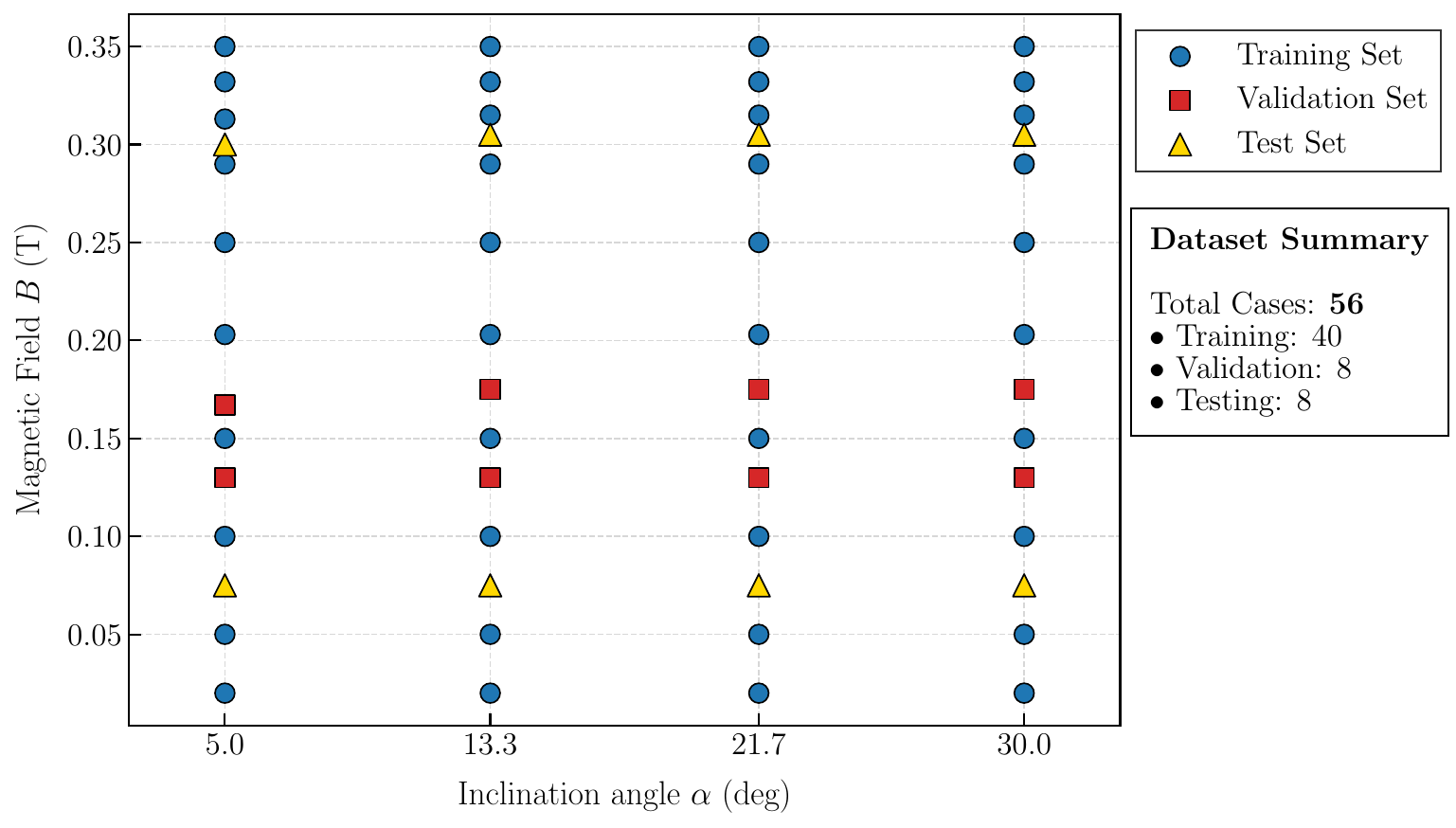}
    \caption{Test, train and validation splitting for a double parametric case.}
    \label{mega_plottino_grid}
\end{figure}

In this section, the more challenging two-parameter analysis is addressed, testing the ability of SHRED to infer flow dynamics for unseen combinations of magnetic field intensity and inclination angle. The \textit{magnetoHDFoam} solver has been used to simulate the metal flow dynamics for 14 different values of the magnetic field intensity and for 4 different inclination angles, for a total of 56 MHD simulations. Specifically, the total set of simulations is graphically represented on a two-dimensional grid in Fig.~\ref{mega_plottino_grid}.
The test sets consist of the complete angular dependencies for two values of the magnetic field intensity, such that both turbulent and laminar regimes are represented. The goal is to test whether SHRED, trained only for different values of the magnetic field with $\alpha \in [5^\circ, 30^\circ]$, is able to infer the full angle-parametric flow dynamics for unseen magnetic field intensities. This task is far from trivial, since the governing MHD equations are highly nonlinear and, therefore, for the same inclination angles, the intensity of the magnetic field can completely determine new dynamics.

\begin{figure}[hbpt]
    \centering
    \includegraphics[width=1\linewidth]{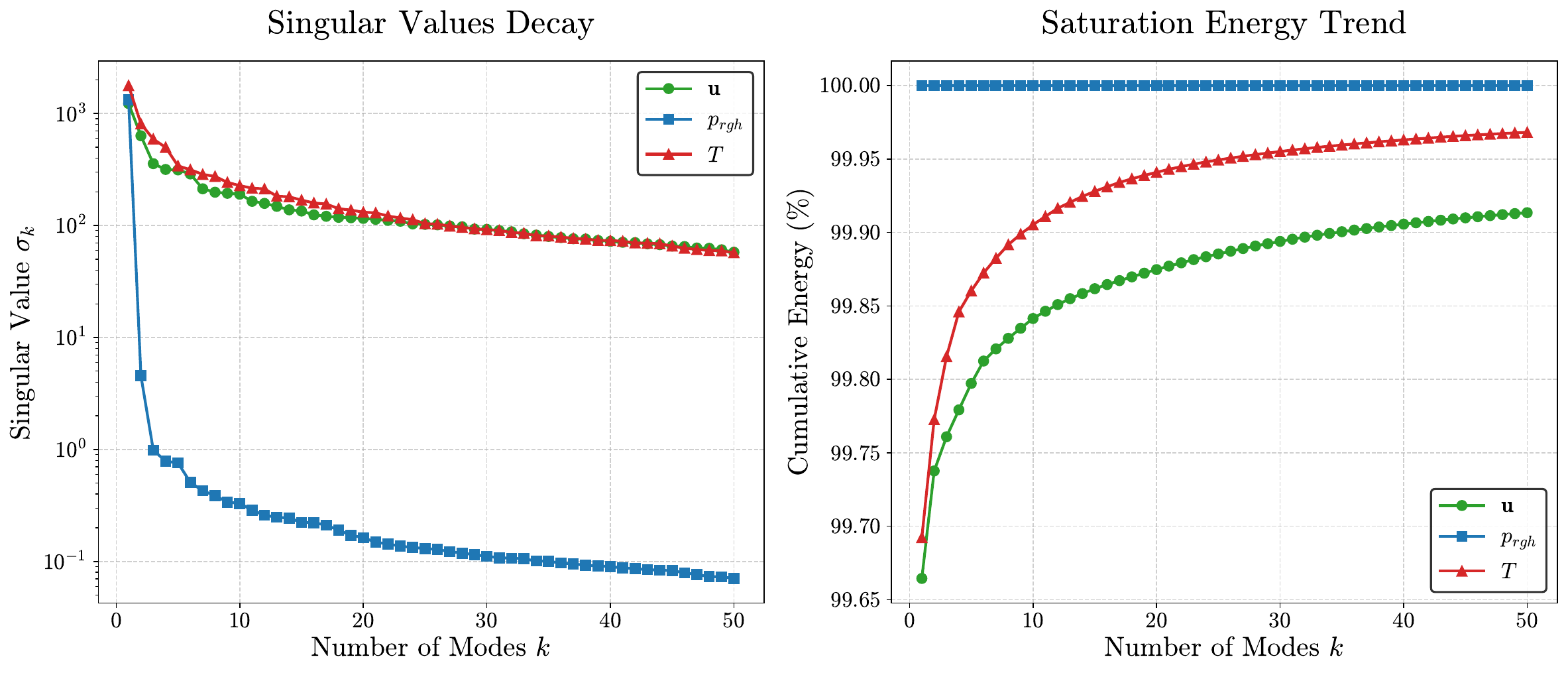}
    \caption{Principal component analysis on training data set for $N_{tr}=40$.}
    \label{mostra_fotografica_1}
\end{figure}

At first, PCA is performed; the PCA coefficients for each field represent the latent dynamics of the temporal evolution of fluctuations around the average snapshot. In Figure ~\ref{mostra_fotografica_1}, on the left the decay of the singular values can be appreciated, while on the right the cumulative energy encoded in the PCA modes, added to the energy of the average field over the total energy, is shown. When compared to the curves of the mono-parametric case, the most evident difference lies in the velocity information encoded in the first PCA coefficient, which embeds 99.65\% of the total energy versus 99.74\% when the magnetic field intensity is the considered parameter. This is coherent with the increasing complexity of the data set as it contains the evolution considering both the angle and intensity as parameters. As in the previous case, with $B_{\text{ext}}$ as the parameter, the first PCA mode for the pressure field, together with the average field, encodes almost the totality of the information. Accordingly, a sufficiently accurate reconstruction can be obtained even if only the average pressure field is considered. This homogenization effect for the pressure field is related, once again, to the Joule damping at high values of the magnetic field in combination with the imposed boundary conditions.

\begin{figure}[htbp]
    \centering
    \includegraphics[width=1\linewidth]{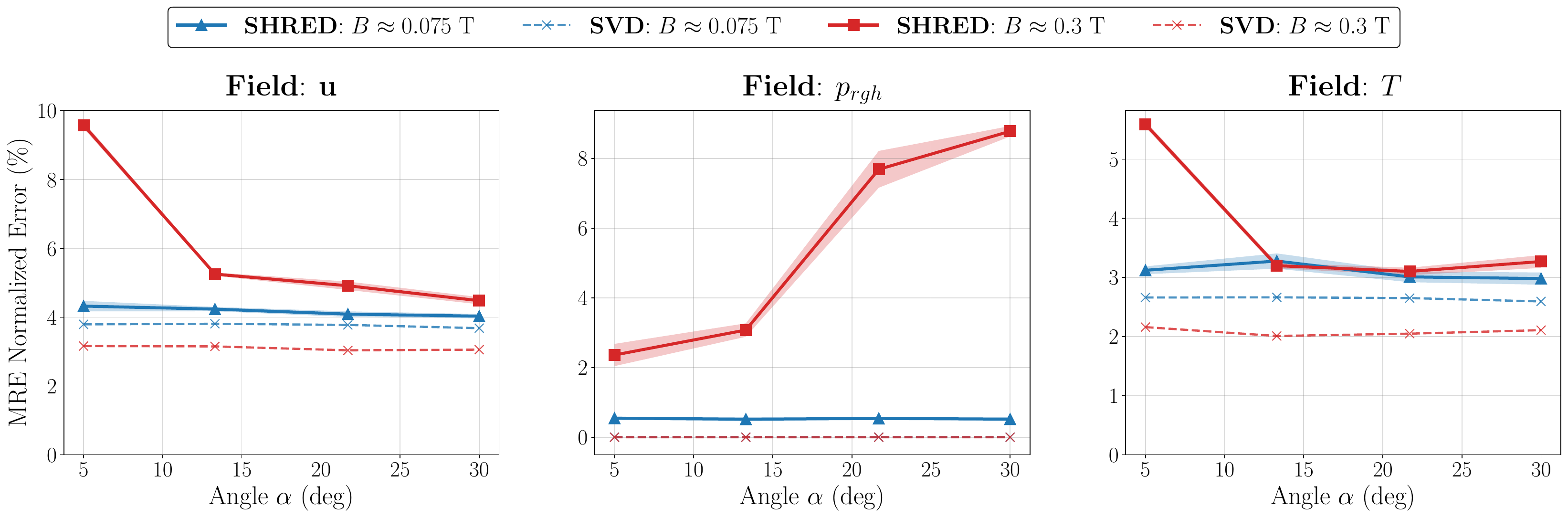}
    \caption{Mean relative error as a function of the inclination angle and as a function of the magnetic field intensity together with the SVD limit.}
    \label{risultati}
\end{figure}

In Figure ~\ref{risultati}, the mean relative error as a function of the inclination angle and as a function of the magnetic field intensity is shown. Moreover, in the same plots, the SVD limit error due to PCA rank truncation is also reported. This lower error bound represents the threshold below which even the most accurate PCA coefficient estimator cannot go, due to the intrinsic truncation associated with the chosen rank and to the fact that the basis functions have not been computed using the test cases. The closer the mean relative error is to the SVD limit (as Principal Component Analysis corresponds to the Singular Value Decomposition of centered data), the closer the SHRED prediction is to the theoretical maximum accuracy achievable within the reduced-order subspace. In particular, it can be seen that in most cases the difference between the SHRED prediction and the error bound is lower than 1\% proving SHRED as a remarkably accurate latent dynamic estimator.

Comparing the error trends of the velocity and temperature fields for an intense magnetic field ($B_{\text{ext}}= 0.3 \ \text{T}$) as a function of the inclination angle, a striking similarity is observed; specifically, for $\alpha=\text{5°}$, the mean relative error is nearly double that of larger angles. This behavior suggests that SHRED effectively learns the strict coupling between the two thermo-hydraulic fields since, at high magnetic field intensities, their relationship is mainly dictated by advection due to the flow laminarization effect. Overall, the mean relative error is almost always lower than 5\%, with some exceptions at high magnetic field intensities; for example, for the pressure field at $\alpha= \{\text{21.7°, 30°}\}$, the error is around 8\%, while for the velocity field at $\alpha= \text{5°}$, it reaches almost 10\%. Despite the remarkable generalization capabilities of SHRED, this is a consequence of the high complexity of the problem, which not only consists of an MHD-modified Hunt flow perturbed by two transversal cylinders whose presence generates vortices and eddies, but also lies in the generalization task itself. Indeed, SHRED is required to assess the dynamics for an unseen magnetic field intensity at specific inclination angles, without the possibility of basing the estimate on measurements coming from cases with the same magnetic field intensity but for a different inclination angle.

\begin{figure}[p]
    \vspace*{\fill}
    \centering
    \includegraphics[width=1\linewidth]{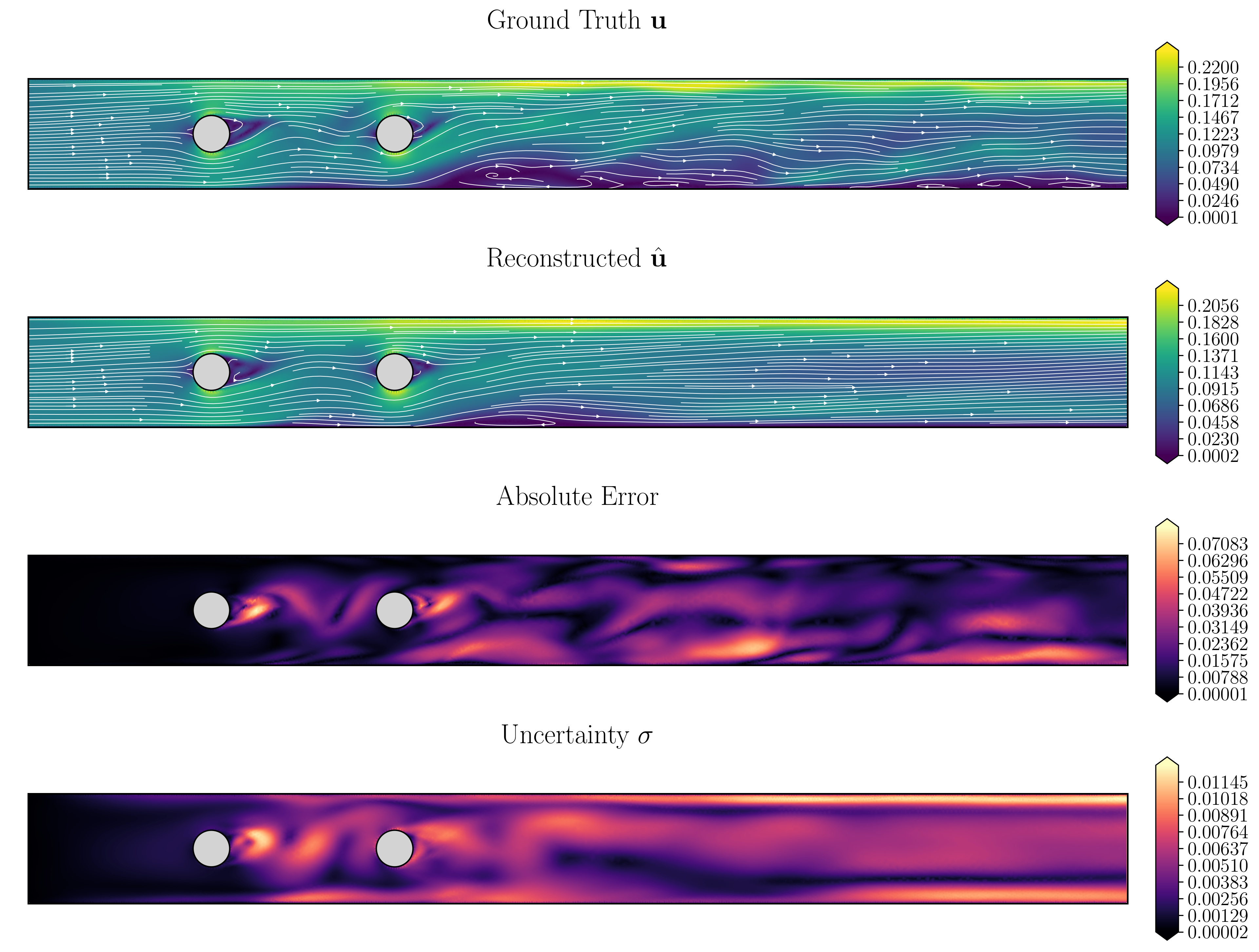}
    \caption{SHRED normalized reconstruction of the velocity field for $|B| = 0.3$ T $\alpha=30$° at $t = 2.5$ s, together with the absolute error and ensemble uncertainty. The neural network effectively captures the formation of a single side layer on the top, accordingly to the specific inclination angle.}
    \label{risultati_f1}
    \vspace*{\fill}
\end{figure}

\begin{figure}[p]
    \vspace*{\fill}
    \centering
    \includegraphics[width=1\linewidth]{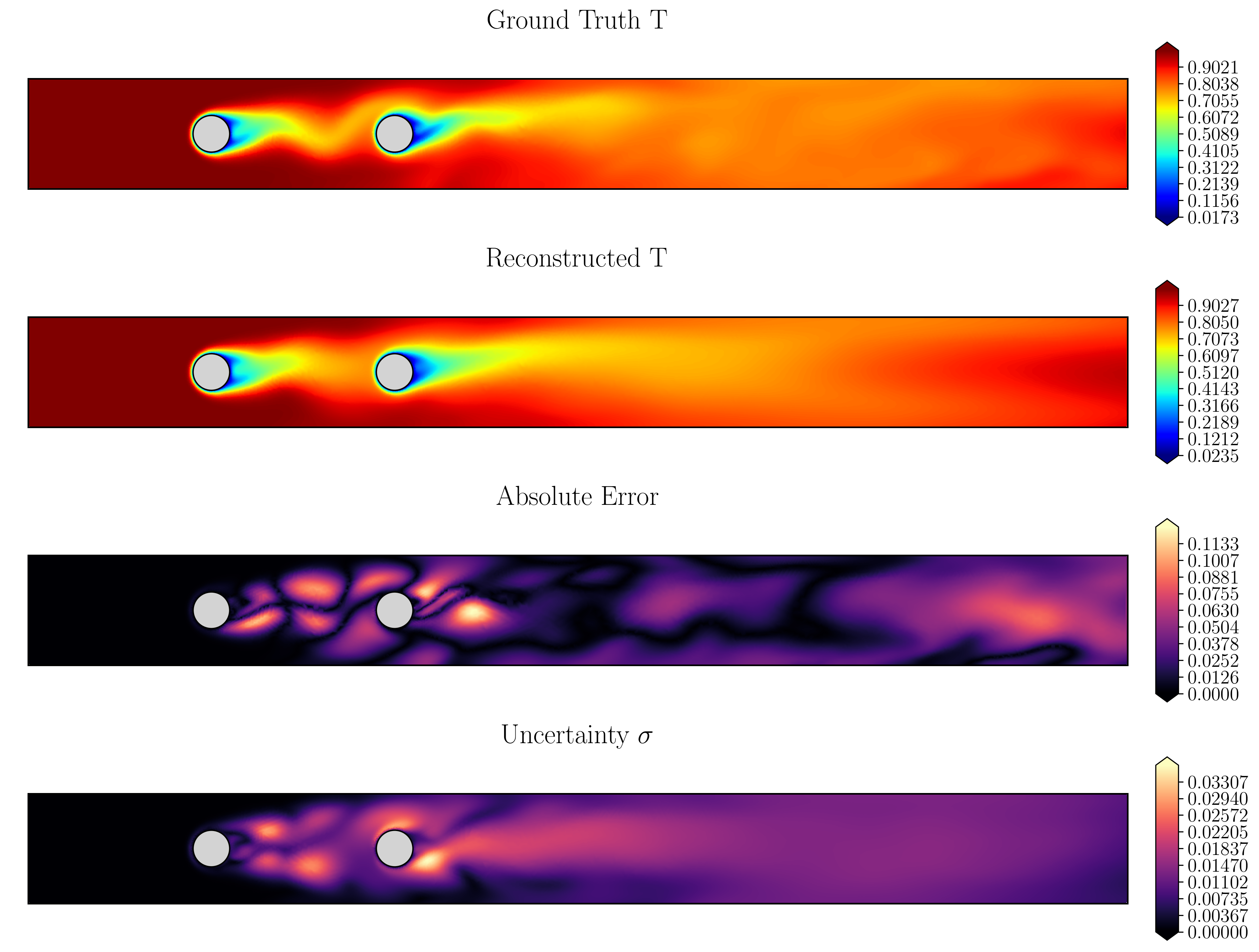}
    \caption{SHRED normalized reconstruction of the temperature field for $|B| = 0.3$ T $\alpha=30$° at $t = 2.5$ s, together with the absolute error and ensemble uncertainty.}
    \label{risultati_f2}
    \vspace*{\fill}
\end{figure}

In conclusion, in Figure ~\ref{risultati_f1} and Figure ~\ref{risultati_f2}, the full-order velocity and temperature fields, together with those estimated by SHRED, for $B_{\text{ext}}=0.3$ T and $\alpha=30^\circ$ are reported. Specifically, the inclination of the magnetic field, no longer parallel to the $z$-axis, dictates an asymmetric formation of the side layer, which occurs exclusively in the upper part of the domain rather than in the lower one. Moreover, the high intensity of the magnetic field is such that laminarization occurs. As previously stated, by visually comparing the velocity streamlines with the temperature field, the advective heat transport can be appreciated, as the hotter regions overall correspond to the regions where velocity streamlines are recirculating, and hence the heat is removed more slowly than where the velocity is higher, as in the side layer. This highlights the strict relationship between velocity streamlines and the temperature field, as dictated by convective heat transport. SHRED reconstruction, as confirmed by the low mean relative error, is capable of predicting these behaviours, confirming its accuracy in performing state estimation. Moreover, an analysis of the residual and uncertainty fields derived from the ensemble strategy reveals a striking similarity in their spatial patterns. This congruence proves the overall reliability of the state estimations, as ensemble SHRED provides, in addition to the estimation of the state, a self-assessment of its own accuracy, demonstrating that regions characterized by higher reconstruction errors inherently coincide with areas where distinct SHRED configurations exhibit localized variance in their predictions.

\clearpage

\clearpage

\clearpage

\section{Conclusions}
\label{concl}

In this work, the application of SHallow Recurrent Decoder (SHRED) in parametric mode to a MHD problem is investigated, where the parameters considered are the inclination angle and the intensity of the externally applied magnetic field, both individually and in combination. A simplified three-dimensional geometry is used to simulate the flow of liquid metal acting as the coolant for the breeding blanket of a Tokamak fusion reactor. The geometry consists of a square duct in which two transversal cylinders are placed in such a way as to simulate the pipes of the heat exchanger, through which water flows to cool down the liquid metal. The overall transversal magnetic field applied, together with the proper boundary conditions, reproduces a generalization of the well-known analytical Hunt flow, as the inclination varies in such a way that an asymmetric formation of side layers is present in addition to the von Kármán shedding generated by the cylinders, which increases the complexity of the reconstruction task.

SHRED learns the non-linear mapping between time measurements of a single observable (e.g., temperature) sampled from sparse sensors (e.g., 4 in a 3D domain) and the latent dynamics of all the thermo-hydraulic fields (temperature, pressure, and velocity), performing indirect state estimation. In its ensemble variation, in addition to the state estimation computed as the average of predictions derived from independent measurements sampled from different sensor positions, SHRED also provides a quantification of the reconstruction uncertainty through the standard deviation fields of each thermo-hydraulic quantity.

Regarding the single-parameter studies, the analysis of SHRED data efficiency with respect to the training set is performed, showing an extraordinary generalization capability: even with only 20\% of the global training set, the mean relative error is always lower than 5\% for each field, considering both the inclination angle and the magnitude as parameters. This is a remarkable result, as SHRED has successfully learned the non-linear relation between the measurements and the latent dynamics as a function of both the angle and intensity parametric dependencies demonstrating its extremely capabilities of generalization, rather than merely fitting data. Moreover, analysing the impact of the DEIM sensor positioning strategy on the reconstruction error, compared to the case in which the sensors are chosen completely at random, the agnosticism of SHRED with respect to sensor arrangement is confirmed, as the performance improvement in the former case, although present, is not particularly marked. The mono-parametric studies proves SHRED data efficiency and SHRED sensor placement agnosticism. 

Regarding the case of two-parameter state estimation in MHD, SHRED reconstruction shows a remarkably low mean relative error, below 5\% for the temperature, pressure, and velocity fields when considering a weak magnetic field ($B_{\text{ext}} = 0.075\ \text{T}$) and inclination angles in the range $[5^\circ, 30^\circ]$. These values are only slightly larger than the lower error bound dictated by the SVD rank truncation, promoting SHRED as a reliable state estimator not only for simple cases but also for more complex and realistic engineering applications in which more than one parametric dependency is inevitably present at the same time. For certain values of the inclination angle at high magnetic field intensities, the mean relative error is slightly larger due to the intrinsic difficulty of the generalization task; nevertheless, even in these challenging cases, SHRED remains sufficiently accurate, with the mean relative error never exceeding 10\%. In addition to the accuracy of SHRED estimations, the possibility to exploit an ensemble of different SHRED models, trained on different sensor configurations, allows for the computation of the prediction trustworthiness, thereby assessing the overall robustness of the estimate. Indeed, the ensemble strategy correctly predicts where higher errors are more likely to occur, thereby enabling a reliable and trustworthy deployment of the model.

In a context of reduced order modelling, as the one adopted in this work being completely data driven, SHRED training requires only few minutes on a standard personal computer; the online stage instead it is almost instantaneous, setting SHRED as real-time and multi-query state estimator for MHD problems. Its data efficiency, sensor agnosticism, noise robustness and computational efficiency, in addition to its accuracy and reliability as a state estimator for complex non-linear MHD mono- or multi-parametric problems, position it as a viable AI-based tool for the online monitoring and control of real Tokamak blanket facilities. Specifically, real-time state estimation can be adopted in synergy with closed-loop feedback control mechanisms, with the ultimate goal of its application within digital-twin frameworks for fusion reactors. In this regard, this work investigated the feasibility of using SHRED as an accurate and reliable state estimator tool paving the way for the deployment of advanced data driven machine learning surrogate models in the next generation of fusion reactor digital twins.
\clearpage
\appendix
\section{MHD ruling equations}
\label{A}
The system of equations solved by \textbf{magnetoHDFoam} is summarized as follows

\begin{itemize}

    \item \textbf{Mass conservation}:
    \begin{equation}
        \frac{\partial \rho}{\partial t} + \nabla \cdot (\rho \mathbf{u}) = 0
    \end{equation}
    with $\rho$ the fluid mass density field and $\mathbf{u}$ the fluid velocity field.
    
    \item \textbf{\textit{Navier-stokes} equation (momentum balance)}:
    \begin{equation}
        \frac{\partial (\rho \mathbf{u})}{\partial t} + \nabla \cdot (\rho \mathbf{u} \mathbf{u}) = -\nabla p_{rgh} + \nabla \cdot \boldsymbol{\tau} + \mathbf{J} \times \mathbf{B}  -\mathbf{g} \cdot \mathbf{h}\nabla\rho
        \label{nav}
    \end{equation}
    with $p_{rgh}=p-\rho g h$, $\boldsymbol{\tau}=\mu (\nabla \mathbf{u}+ (\nabla \mathbf{u})^T)$ is the viscous stress tensor for an incompressible flow ($\nabla \cdot\mathbf{u}=0$), $\mathbf{J} \times \mathbf{B}$ is the Lorentz force and $\mathbf{g} \cdot \mathbf{h}\nabla\rho$ are the \textit{buoyancy} forces, where $\mathbf{g}$ is the gravitational acceleration and $\mathbf{h}$ is the vertical height.
    
    \item \textbf{Induction equation}:
    \begin{equation}
       \frac{\partial \mathbf{B}}{\partial t}= \nabla \times (\mathbf{u \times \mathbf{B}})+ \eta_m \nabla^2 \mathbf{B}
    \end{equation}
    with the constraint of a solenoidal magnetic field $\nabla \cdot \mathbf{B} = 0$ that must be ensured.
    
    \item \textbf{Internal energy balance}:
    \begin{equation}
        \frac{\partial (\rho e)}{\partial t} + \nabla \cdot (\rho \mathbf{u} e) = \nabla \cdot (\alpha_{\text{eff}} \nabla e) + \frac{|\mathbf{J}|^2}{\sigma}
    \end{equation}
    where $e$ is the internal energy and $\alpha_{\text{eff}}$ is the effective thermal diffusivity.

    \item \textbf{State equation}:
    \begin{equation}
        \rho=\rho(T,p)
    \end{equation}
    which expresses the link between the density of the fluid with the pressure and temperature. Physically, the fluid under investigation (e.g., liquid metal) is incompressible. However, to model buoyancy forces (natural convection), the density cannot be treated as a constant but it must vary as a function of temperature. Therefore, the solver treats the density as a variable dominated by thermal expansion effects ($\frac{\partial \rho}{\partial T}$), while the dependence on pressure ($\frac{\partial \rho}{\partial p}$, related to acoustic compressibility) remains negligible (Boussinesq approximation \cite{incropera_fundamentals_2011,INTROINI2023112118}).
\end{itemize}

\bibliographystyle{unsrt}  
\bibliography{references}

@InProceedings{10.1007/BFb0091924,
author="Takens, Floris",
editor="Rand, David
and Young, Lai-Sang",
title="Detecting strange attractors in turbulence",
booktitle="Dynamical Systems and Turbulence, Warwick 1980",
year="1981",
publisher="Springer Berlin Heidelberg",
address="Berlin, Heidelberg",
pages="366--381",
isbn="978-3-540-38945-3"
}

@book{quarteroni_reduced_2015,
	edition = {1},
	series = {{UNITEXT}},
	title = {Reduced {Basis} {Methods} for {Partial} {Differential} {Equations}: {An} {Introduction}},
	isbn = {978-3-319-15431-2},
	url = {https://doi.org/10.1007/978-3-319-15431-2},
	publisher = {Springer Cham},
	author = {Quarteroni, A and Manzoni, A and Negri, F},
	year = {2015},
}

@book{brunton_data-driven_2022,
	address = {USA},
	edition = {2nd},
	title = {Data-{Driven} {Science} and {Engineering}: {Machine} {Learning}, {Dynamical} {Systems}, and {Control}},
	isbn = {1-00-909848-9},
	publisher = {Cambridge University Press},
	author = {Brunton, Steven L and Kutz, J Nathan},
	year = {2022},
    url={https://www.cambridge.org/core/books/datadriven-science-and-engineering/77D52B171B60A496EAFE4DB662ADC36E}
}

@article{riva_hybrid_2023-1,
	title = {Hybrid data assimilation methods, {Part} {I}: {Numerical} {comparison} between {GEIM} and {PBDW}},
	volume = {190},
	copyright = {All rights reserved},
	issn = {0306-4549},
	url = {https://www.sciencedirect.com/science/article/pii/S0306454923001834},
	doi = {https://doi.org/10.1016/j.anucene.2023.109864},
	journal = {Annals of Nuclear Energy},
	author = {Riva, Stefano and Introini, Carolina and Lorenzi, Stefano and Cammi, Antonio},
	year = {2023},
	pages = {109864},
}

@article{williams_sensing_2023,
author = {Williams, Jan P. and Zahn, Olivia and Kutz, J. Nathan},
title = {Sensing with shallow recurrent decoder networks},
journal = {Proceedings of the Royal Society A: Mathematical, Physical and Engineering Sciences},
volume = {480},
number = {2298},
pages = {20240054},
year = {2024},
doi = {10.1098/rspa.2024.0054},
url = {https://royalsocietypublishing.org/doi/abs/10.1098/rspa.2024.0054}
}

@article{ebers_leveraging_2023,
	author = {Ebers, Megan R. and Williams, Jan P. and Steele, Katherine M. and Nathan Kutz, J.},
	journal = {IEEE Access},
	title = {Leveraging Arbitrary Mobile Sensor Trajectories With Shallow Recurrent Decoder Networks for Full-State Reconstruction},
	year = {2024},
	volume = {12},
	pages = {97428-97439},
	doi = {10.1109/ACCESS.2024.3423679},
	url = {https://doi.org/10.1109/ACCESS.2024.3423679}
}

@article{kutz_shallow_2024,
	doi = {10.1088/2632-2153/adcd20},
	url = {https://dx.doi.org/10.1088/2632-2153/adcd20},
	year = {2025},
	month = {apr},
	publisher = {IOP Publishing},
	volume = {6},
	number = {2},
	pages = {025024},
	author = {Faraji, Farbod and Reza, Maryam and Kutz, J Nathan},
	title = {Shallow recurrent decoder for reduced order modeling of E × B plasma dynamics},
	journal = {Machine Learning: Science and Technology}
}

@article{riva2024robuststateestimationpartial,
title = {Robust state estimation from partial out-core measurements with Shallow Recurrent Decoder for nuclear reactors},
journal = {Progress in Nuclear Energy},
volume = {189},
pages = {105928},
year = {2025},
issn = {0149-1970},
doi = {https://doi.org/10.1016/j.pnucene.2025.105928},
url = {https://www.sciencedirect.com/science/article/pii/S0149197025003269},
author = {Stefano Riva and Carolina Introini and Antonio Cammi and J. Nathan Kutz}
}

@article{shredrom,	
	title = {Reduced order modeling with shallow recurrent decoder networks},
  volume = {16},
  ISSN = {2041-1723},
  url = {http://dx.doi.org/10.1038/s41467-025-65126-y},
  DOI = {10.1038/s41467-025-65126-y},
  number = {1},
  journal = {Nature Communications},
  publisher = {Springer Science and Business Media LLC},
  author = {Tomasetto, Matteo and Williams, Jan P. and Braghin, Francesco and Manzoni, Andrea and Kutz, J. Nathan},
  year = {2025},
  month = nov 
}

@article{hochreiter1997long,
    author = {Hochreiter, Sepp and Schmidhuber, Jürgen},
    title = {Long Short-Term Memory},
    journal = {Neural Computation},
    volume = {9},
    number = {8},
    pages = {1735-1780},
    year = {1997},
    month = {11},
    doi = {10.1162/neco.1997.9.8.1735},
    url = {https://doi.org/10.1162/neco.1997.9.8.1735}
}

@article{erichson_shallow_2020,
	title = {Shallow neural networks for fluid flow reconstruction with limited sensors},
	volume = {476},
	issn = {1364-5021, 1471-2946},
	url = {https://royalsocietypublishing.org/doi/10.1098/rspa.2020.0097},
	doi = {10.1098/rspa.2020.0097},
	number = {2238},
	journal = {Proceedings of the Royal Society A: Mathematical, Physical and Engineering Sciences},
	author = {Erichson, N. Benjamin and Mathelin, Lionel and Yao, Zhewei and Brunton, Steven L. and Mahoney, Michael W. and Kutz, J. Nathan},
	month = jun,
	year = {2020},
	pages = {20200097},
}

@article{HORNIK1989359,
title = {Multilayer feedforward networks are universal approximators},
journal = {Neural Networks},
volume = {2},
number = {5},
pages = {359-366},
year = {1989},
issn = {0893-6080},
doi = {https://doi.org/10.1016/0893-6080(89)90020-8},
url = {https://www.sciencedirect.com/science/article/pii/0893608089900208},
author = {Kurt Hornik and Maxwell Stinchcombe and Halbert White}
}

@article{Chen1995_UniversalOperator,
	author = {Chen, Tianping and Chen, Hong},
	title = {Universal approximation to nonlinear operators by neural networks with arbitrary activation functions and its application to dynamical systems},
	journal = {IEEE Transactions on Neural Networks},
	year = {1995},
	volume = {6},
	number = {4},
	pages = {911--917},
	doi = {10.1109/72.392253},
	issn = {1045-9227},
	note = {PMID: 18263379}
}

@article{MARTELLI2019183,
title = {Literature review of lead-lithium thermophysical properties},
journal = {Fusion Engineering and Design},
volume = {138},
pages = {183-195},
year = {2019},
issn = {0920-3796},
doi = {https://doi.org/10.1016/j.fusengdes.2018.11.028},
url = {https://www.sciencedirect.com/science/article/pii/S0920379618307361},
author = {D. Martelli and A. Venturini and M. Utili}
}

@article{article,
author = {Frank, Markus and Barleon, L. and Müller, U.},
year = {2001},
month = {08},
pages = {2287-2295},
title = {Visual analysis of two-dimensional magnetohydrodynamics},
volume = {13},
journal = {Physics of Fluids - PHYS FLUIDS},
doi = {10.1063/1.1383785}
}

@unpublished{loverso2025reduced,
  title={Reduced Order Modeling for nuclear fusion with parametric Shallow Recurrent Decoder Networks: Applications to Magnetohydrodynamics},
  author={Lo Verso, Matteo and Cammi, Antonio and Kutz, J. Nathan},
  note={Submitted to Physics of Fluids},
  year={2025}
}

@misc{riva2025towards,
      title={Towards Efficient Parametric State Estimation in Circulating Fuel Reactors with Shallow Recurrent Decoder Networks}, 
      author={Stefano Riva and Carolina Introini and J. Nathan Kutz and Antonio Cammi},
      year={2025},
      eprint={2503.08904},
      archivePrefix={arXiv},
      primaryClass={cs.LG},
      url={https://arxiv.org/abs/2503.08904}, 
      note={under review at Chemical Engineering Science}
}

@book{bird2006transport,
  title={Transport Phenomena},
  author={Bird, R. Byron and Stewart, Warren E. and Lightfoot, Edwin N.},
  edition={Revised 2nd},
  year={2006},
  publisher={John Wiley \& Sons}
}

@incollection{buhler2007liquid,
  title={Liquid metal magnetohydrodynamics for fusion blankets},
  author={Molokov, S. and Moreau, R. and Moffatt, K. and B{\"u}hler, L.},
  booktitle={Magnetohydrodynamics: Historical Evolution and Trends},
  pages={171--194},
  year={2007},
  publisher={Springer Dordrecht}
}

@article{RevModPhys.76.1071,
  title = {Physics of magnetically confined plasmas},
  author = {Boozer, Allen H.},
  journal = {Rev. Mod. Phys.},
  volume = {76},
  issue = {4},
  pages = {1071--1141},
  numpages = {0},
  year = {2005},
  month = {Jan},
  publisher = {American Physical Society},
  doi = {10.1103/RevModPhys.76.1071},
  url = {https://link.aps.org/doi/10.1103/RevModPhys.76.1071}
}

@article{unknown1,
author = {Tassone, Alessandro},
year = {2022},
month = {02},
pages = {},
title = {On the magnetic field distribution in the TBM set and a blanket based on the DEMO2017 baseline},
doi = {10.13140/RG.2.2.34998.72004}
}

@book{davidson2015turbulence,
  title={Turbulence: An Introduction for Scientists and Engineers},
  author={Davidson, Peter A.},
  edition={2nd},
  year={2015},
  publisher={Oxford University Press}
}

@article{riva_multi-physics_2024,
	title = {Multi-physics model bias correction with data-driven reduced order techniques: {Application} to nuclear case studies},
	volume = {135},
	copyright = {All rights reserved},
	issn = {0307-904X},
	url = {https://www.sciencedirect.com/science/article/pii/S0307904X24003196},
	doi = {https://doi.org/10.1016/j.apm.2024.06.040},
	journal = {Applied Mathematical Modelling},
	author = {Riva, Stefano and Introini, Carolina and Cammi, Antonio},
	year = {2024},
	pages = {243--268}
}

@inproceedings{loverso2024novel,
  title={A novel Openfoam library for magneto-hydrodynamics studies in the nuclear fusion field},
  author={Lo Verso, Matteo and Introini, C. and Cervi, E. and Barucca, L. and Caramello, M. and Di Prinzio, M. and Giacobbo, F. and Savoldi, L. and Cammi, A.},
  booktitle={Proceedings of the NUTHOS-14 International Conference},
  address={Vancouver, Canada},
  year={2024}
}

@inproceedings{loverso2025magnetohdfoam,
  title={magnetoHDFoam: a novel open-source OpenFOAM library for magnetohydrodynamics},
  author={Lo Verso, Matteo and Introini, C. and Cammi, A.},
  booktitle={Open Source Software for Fusion Energy (OSSFE) Conference},
  year={2025}
}

@article{INTROINI2023112118,
	title = {A complete {CFD} study on natural convection in the {TRIGA} {Mark} {II} reactor},
	journal = {Nuclear Engineering and Design},
	volume = {403},
	pages = {112118},
	year = {2023},
	issn = {0029-5493},
	doi = {10.1016/j.nucengdes.2022.112118},
	url = {https://www.sciencedirect.com/science/article/pii/S0029549322004691},
	author = {Carolina Introini and Davide Chiesa and Massimiliano Nastasi and Ezio Previtali and Andrea Salvini and Monica Sisti and Xiang Wang and Antonio Cammi}
}

@book{incropera_fundamentals_2011,
	edition = {7},
	title = {Fundamentals of {Heat} and {Mass} {Transfer}},
	publisher = {John Wiley \& Sons, Ltd},
	author = {Incropera, Frank P and Bergman, Theodore L and Lavine, Adrienne S and Dewitt, David P},
	year = {2011}
}

@book
{
  greenshieldsweller2022,
  title     = "Notes on Computational Fluid Dynamics: General Principles",
  author    = "Greenshields, Christopher and Weller, Henry",
  year      = 2022,
  publisher = "CFD Direct Ltd",
  address   = "Reading, UK"
}

@misc{riva2025_parametricMSFR,
	title = {Towards {Efficient} {Parametric} {State} {Estimation} in {Circulating} {Fuel} {Reactors} with {Shallow} {Recurrent} {Decoder} {Networks}},
	url = {http://arxiv.org/abs/2503.08904},
	doi = {10.48550/arXiv.2503.08904},
	publisher = {arXiv},
	author = {Riva, Stefano and Introini, Carolina and Kutz, J. Nathan and Cammi, Antonio},
	month = mar,
	year = {2025},
	note = {arXiv:2503.08904 [cs]}
}

@book{wesson2011tokamaks,
  title={Tokamaks},
  author={Wesson, J. and Campbell, D.J.},
  isbn={9780199592234},
  lccn={2012359121},
  series={International Series of Monographs on Physics},
  url={https://books.google.it/books?id=XJssMXjHUr0C},
  year={2011},
  publisher={OUP Oxford}
}

@book{biskamp1997nonlinear,
  title={Nonlinear Magnetohydrodynamics},
  author={Biskamp, D. and Biskamp, D.},
  isbn={9780521599184},
  lccn={97201462},
  series={Cambridge Monographs on Plasma Physics},
  url={https://books.google.it/books?id=OzFNhaVKA48C},
  year={1997},
  publisher={Cambridge University Press}
}

@article{TheDEMO,
author = {Arena, Pietro and Del Nevo, Alessandro and Moro, F. and Noce, Simone and Mozzillo, Rocco and Imbriani, Vito and Giannetti, Fabio and Edemetti, Francesco and Froio, Antonio and Savoldi, Laura and Siriano, Simone and Tassone, Alessandro and Urgorri, Fernando and Maio, Pietro and Catanzaro, Ilenia and Bongiovì, Gaetano},
year = {2021},
month = {12},
pages = {11592},
title = {The DEMO Water-Cooled Lead–Lithium Breeding Blanket: Design Status at the End of the Pre-Conceptual Design Phase},
volume = {11},
journal = {Applied Sciences},
doi = {10.3390/app112411592}
}

@article{SIRIANO2025126840,
title = {Numerical investigation of liquid metal magneto-convection at high Grashof and Hartmann number in a prototypical water-cooled breeding blanket for fusion reactors},
journal = {International Journal of Heat and Mass Transfer},
volume = {242},
pages = {126840},
year = {2025},
issn = {0017-9310},
doi = {https://doi.org/10.1016/j.ijheatmasstransfer.2025.126840},
url = {https://www.sciencedirect.com/science/article/pii/S0017931025001814},
author = {Simone Siriano and Alessandro Tassone and Lorenzo Melchiorri and Gianfranco Caruso}
}

@article{MARTELLI201848,
title = {Thermo-hydraulic analysis of EU DEMO WCLL breeding blanket},
journal = {Fusion Engineering and Design},
volume = {130},
pages = {48-55},
year = {2018},
issn = {0920-3796},
doi = {https://doi.org/10.1016/j.fusengdes.2018.03.030},
url = {https://www.sciencedirect.com/science/article/pii/S0920379618302400},
author = {Emanuela Martelli and Gianfranco Caruso and Fabio Giannetti and Alessandro {Del Nevo}}
}

@misc{riva2025constrainedsensingreliablestate,
      title={Constrained Sensing and Reliable State Estimation with Shallow Recurrent Decoders on a TRIGA Mark II Reactor}, 
      author={Stefano Riva and Carolina Introini and Josè Nathan Kutz and Antonio Cammi},
      year={2025},
      eprint={2510.12368},
      archivePrefix={arXiv},
      primaryClass={cs.CE},
      url={https://arxiv.org/abs/2510.12368}, 
}

@unknown{From_Models_To-Experiments,
author = {Riva, Stefano and Missaglia, Andrea and Introini, Carolina and Kutz, J. and Cammi, Antonio},
year = {2026},
month = {04},
pages = {},
title = {From Models To Experiments: Shallow Recurrent Decoder Networks on the DYNASTY Experimental Facility},
doi = {10.48550/arXiv.2503.08907}
}

@misc{verso2026applicationparametricshallowrecurrent,
      title={Application of parametric Shallow Recurrent Decoder Network to magnetohydrodynamic flows in liquid metal blankets of fusion reactors}, 
      author={M. Lo Verso and C. Introini and E. Cervi and L. Savoldi and J. N. Kutz and A. Cammi},
      year={2026},
      eprint={2604.02139},
      archivePrefix={arXiv},
      primaryClass={cs.LG},
      url={https://arxiv.org/abs/2604.02139}, 
}

@misc{tomasetto2025reducedordermodelingshallow,
      title={Reduced Order Modeling with Shallow Recurrent Decoder Networks}, 
      author={Matteo Tomasetto and Jan P. Williams and Francesco Braghin and Andrea Manzoni and J. Nathan Kutz},
      year={2025},
      eprint={2502.10930},
      archivePrefix={arXiv},
      primaryClass={cs.LG},
      url={https://arxiv.org/abs/2502.10930}, 
}

@article{Jalalvand,
author = {Jalalvand, Azarakhsh and Abbate, Joseph and Conlin, Rory and Verdoolaege, Geert and Kolemen, Egemen},
year = {2021},
month = {06},
pages = {1-12},
title = {Real-Time and Adaptive Reservoir Computing With Application to Profile Prediction in Fusion Plasma},
volume = {PP},
journal = {IEEE Transactions on Neural Networks and Learning Systems},
doi = {10.1109/TNNLS.2021.3085504}
}

@article{Seo,
author = {Seo, Jaemin and Kim, SangKyeun and Jalalvand, Azarakhsh and Conlin, Rory and Rothstein, Andrew and Abbate, Joseph and Erickson, Keith and Wai, Josiah and Shousha, Ricardo and Kolemen, Egemen},
year = {2024},
month = {02},
pages = {746-751},
title = {Avoiding fusion plasma tearing instability with deep reinforcement learning},
volume = {626},
journal = {Nature},
doi = {10.1038/s41586-024-07024-9}
}

@article{Degrave,
author = {Degrave, Jonas and Felici, Federico and Buchli, Jonas and Neunert, Michael and Tracey, Brendan and Carpanese, Francesco and Ewalds, Timo and Hafner, Roland and Abdolmaleki, Abbas and Casas, Diego and Donner, Craig and Fritz, Leslie and Galperti, Cristian and Huber, Andrea and Keeling, James and Tsimpoukelli, Maria and Kay, Jackie and Merle, Antoine and Moret, Jean-Marc and Riedmiller, Martin},
year = {2022},
month = {02},
pages = {414-419},
title = {Magnetic control of tokamak plasmas through deep reinforcement learning},
volume = {602},
journal = {Nature},
doi = {10.1038/s41586-021-04301-9}
}

@mastersthesis{CS,
  author  = {Scardino, Claudio},
  title   = {State Estimation in {M}agneto-hydrodynamics: from interpolation to recurrent networks},
  school  = {Politecnico di Milano, Nuclear Engineering},
  year    = {2026},
  address = {Milan, Italy},
  month   = {March}
}

@article{Gong03062022,
author = {Helin Gong and Sibo Cheng and Zhang Chen and Qing Li},
title = {Data-Enabled Physics-Informed Machine Learning for Reduced-Order Modeling Digital Twin: Application to Nuclear Reactor Physics},
journal = {Nuclear Science and Engineering},
volume = {196},
number = {6},
pages = {668--693},
year = {2022},
publisher = {Taylor \& Francis},
doi = {10.1080/00295639.2021.2014752},


URL = { 
    
        https://doi.org/10.1080/00295639.2021.2014752
    
    

},
eprint = { 
    
        https://doi.org/10.1080/00295639.2021.2014752
    
    

}

}

@article{GONG2022109431,
title = {An efficient digital twin based on machine learning SVD autoencoder and generalised latent assimilation for nuclear reactor physics},
journal = {Annals of Nuclear Energy},
volume = {179},
pages = {109431},
year = {2022},
issn = {0306-4549},
doi = {https://doi.org/10.1016/j.anucene.2022.109431},
url = {https://www.sciencedirect.com/science/article/pii/S0306454922004613},
author = {Helin Gong and Sibo Cheng and Zhang Chen and Qing Li and César Quilodrán-Casas and Dunhui Xiao and Rossella Arcucci}
}

@article{leite_application_2025,
  author  = {Leite, Victor Coppo and Merzari, Elia and Novak, April
             and Ponciroli, Roberto and Ibarra, Lander},
  title   = {Application of a Physics-Informed Convolutional Neural
             Network for Monitoring the Temperature Fields in
             High-Temperature Gas Reactors},
  journal = {Nuclear Science and Engineering},
  volume  = {199},
  number  = {10},
  pages   = {1712--1732},
  year    = {2025},
  doi     = {10.1080/00295639.2024.2443337}
}

@article{gong_reactor_2024,
  author  = {Gong, He-Lin and Li, Han and Xiao, Dunhui and Cheng, Sibo},
  title   = {Reactor Field Reconstruction from Sparse and Movable
             Sensors Using {Voronoi} Tessellation-Assisted
             Convolutional Neural Networks},
  journal = {Nuclear Science and Techniques},
  volume  = {35},
  pages   = {43},
  year    = {2024},
  doi     = {10.1007/s41365-024-01400-w}
}

@article{YANG2023109656,
title = {A data-enabled physics-informed neural network with comprehensive numerical study on solving neutron diffusion eigenvalue problems},
journal = {Annals of Nuclear Energy},
volume = {183},
pages = {109656},
year = {2023},
issn = {0306-4549},
doi = {https://doi.org/10.1016/j.anucene.2022.109656},
url = {https://www.sciencedirect.com/science/article/pii/S0306454922006867},
author = {Yu Yang and Helin Gong and Shiquan Zhang and Qihong Yang and Zhang Chen and Qiaolin He and Qing Li}
}

\end{document}